\documentclass[%
 reprint,
 amsmath,amssymb,
 aps,
]{revtex4-2}

\usepackage{graphicx}
\usepackage{dcolumn}
\usepackage{bm}

\usepackage{nomencl}
\usepackage{siunitx}
\makenomenclature
\usepackage{float}
\usepackage{booktabs}
\usepackage{etoolbox}
\renewcommand\nomgroup[1]{%
  \item[\bfseries
  \ifstrequal{#1}{A}{Acronyms}{%
  \ifstrequal{#1}{B}{Electromagnetics}{%
  \ifstrequal{#1}{C}{Matter/Electrons}{
  \ifstrequal{#1}{D}{Weyl and Dirac material properties}{}}}}%
]}

\begin{document}


\title{Plasmon Fizeau drag in 3D Dirac and Weyl semimetals}

\author{Morgan G. Blevins}
\affiliation{Department of Electrical Engineering and Computer Science, Massachusetts Institute of Technology, Cambridge, MA 02139, USA}
\affiliation{Draper Scholar, Draper, Cambridge, MA 02139, USA}

\author{Svetlana V. Boriskina}%
 \email{sborisk@mit.edu}
\affiliation{%
 Department of Mechanical Engineering, Massachusetts Institute of Technology, Cambridge, MA 02139, USA}%


\date{\today}

\begin{abstract}
There is a need for compact, dynamically tunable nonreciprocal optical elements to enable on-chip-compatible optical isolators and more efficient radiative energy transfer systems.
Plasmon Fizeau drag, the drag of electrical current on propagating surface plasmon polaritons, has been proposed to induce nonreciprocal surface modes to enable one-way energy transport. 
However, relativistic electron drift velocities are required to induce appreciable contrast between the dispersion characteristics of co-propagating and counter-propagating surface plasmon modes. 
The high electron drift velocity of graphene previously allowed for the experimental demonstration of current-induced nonreciprocity in a two-dimensional (2D) Dirac material. 
The high electron drift and Fermi velocities in three-dimensional (3D) Dirac materials make them ideal candidates for the effect, however, both the theory of the Fizeau drag effect and its experimental demonstrations in 3D Dirac materials are missing. 
Here we develop a comprehensive theory of Fizeau drag in DC-biased 3D Weyl semimetals (WSM) or Dirac semimetals (DSM), both under local and non-local approximation and with dissipative losses. 
We predict that under practical assumptions for loss, Fizeau drag in the DSM Cd$_3$As$_2$ opens windows of pseudo-unidirectional transport. 
We additionally introduce new figures of merit to rank nonreciprocal plasmonic systems by their potential for directional SPP transport. 
Further, we propose a new approach for achieving appreciable plasmonic Fizeau drag via optically pumping bulk inversion symmetry breaking WSMs or DSMs. 

\end{abstract}

\maketitle


\section*{Introduction}\label{sec1}
The inherent reciprocity of conventional optical materials imposes fundamental limits on the efficiency and performance of optical systems such as lasers and radiative energy harvesters\cite{Boriskina2022TheLight}. Nonreciprocal optical elements prevent bi-directional wave propagation, allowing propagation of a forward-moving optical mode while preventing the corresponding backward mode propagation, and thus enabling the development of optical isolators and circulators\cite{Caloz2018ElectromagneticNonreciprocity,Jalas2013WhatIsolator}. While magneto-optic materials make up conventional nonreciprocal devices, there is a growing interest in devices that do not require external magnetic fields, which can be incompatible with other device materials and require bulky designs\cite{Bi2011On-chipResonators}. Alternative approaches via space-time modulation to the material permittivity\cite{Guo2019NonreciprocalModulation, Sounas2017Non-reciprocalModulation, Fan2018NonreciprocalMagneto-optics} and nonlinear optical effects\cite{Fan_Wang_Varghese_Shen_Niu_Xuan_Weiner_Qi_2012, Bender_Observation_Nonlinearities2013, Chang_Jiang_Hua_Yang_Wen_Jiang_Li_Wang_Xiao_2014, Mahmoud_Davoyan_Engheta_2015} allow for on-chip integration, but are severely limited by the power consumption requirements\cite{Khurgin_2023}.

Another emerging alternative for nonreciprocal optical devices, which requires neither an external magnetic field nor ultra-fast modulation, is based on making use of the excitation of surface plasmon polariton (SPP) modes at the interfaces of Weyl semimetals (WSM)\cite{Kotov2019GiantSemimetals, Hofmann2016SurfaceSemimetals,Zhao2020Axion-Field-EnabledSemimetals, Tsurimaki_Qian_Pajovic_Han_Li_Chen_2020, Pajovic2020IntrinsicSurfaces, Guo2023LightSemimetals, Tang_Chen_Zhang_2021}. WSMs are topological materials with distinct electronic bandstructure, which supports pairs of non-degenerate, chiral Weyl nodes formed from intersecting, linearly dispersing bands. WSM must be either inversion symmetry (I) or time-reversal symmetry (TRS) breaking, or both, to achieve the non-degenerate bandstructure. In TRS-breaking WSMs, the flux of Berry curvature between Weyl nodes of opposite chirality plays a role of a pseudo-magnetic field in the momentum space, giving rise to the anomalous Hall effect and off-diagonal components in bulk dielectric tensors of these materials. As a result, they act like nonreciprocal magneto-optical materials without the need for an external magnetic field, and have been studied for applications in tunable near-field radiative heat transfer\cite{Tsurimaki_Qian_Pajovic_Han_Li_Chen_2020, Pajovic2020IntrinsicSurfaces, Zhao2020Axion-Field-EnabledSemimetals} and as optical isolator material candidates\cite{Asadchy2020Sub-WavelengthSemimetals,Chistyakov2023TunableSemimetals}.

\begin{figure*} 
    \centering
    \includegraphics[width=12.9cm]{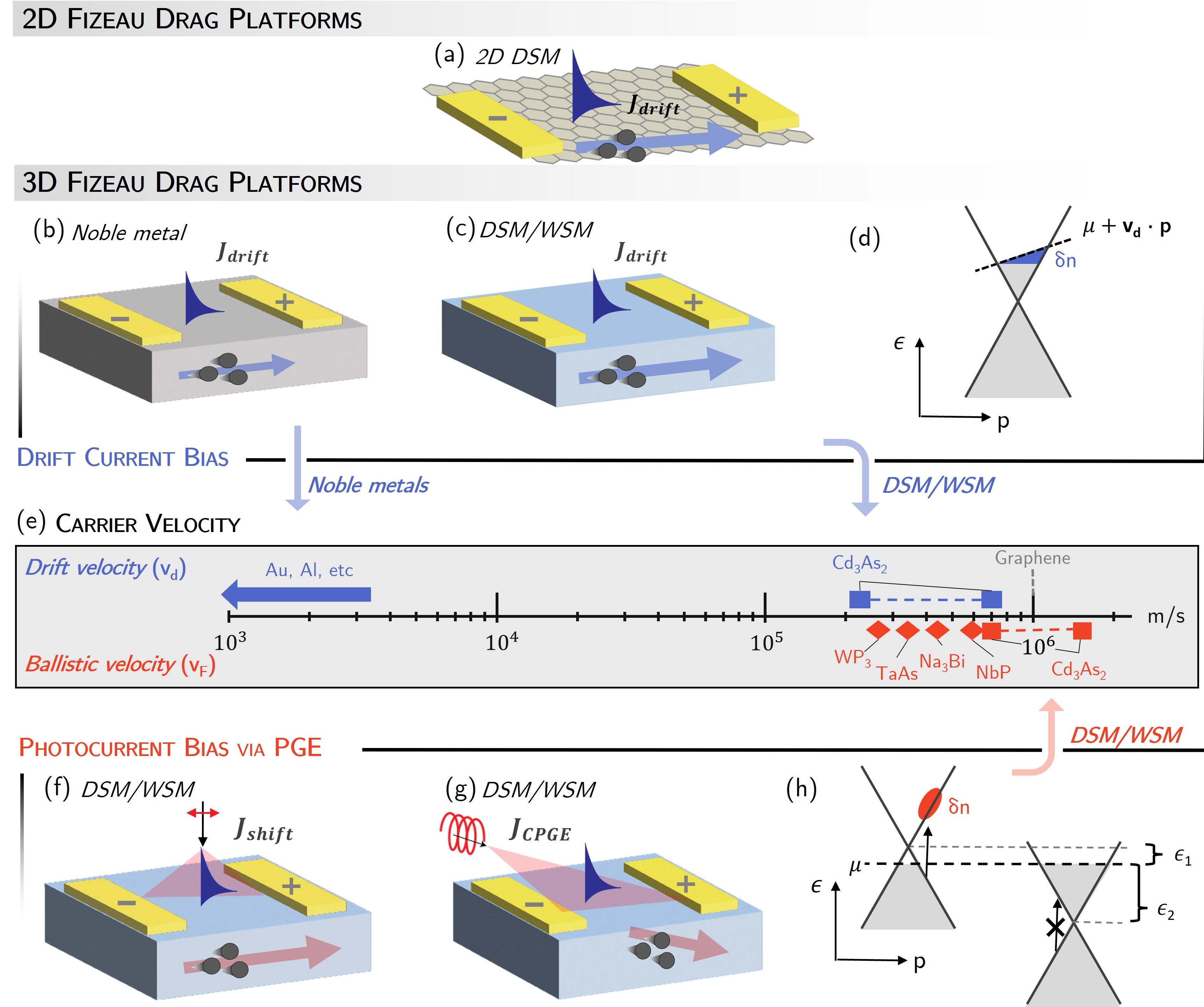}
    \caption{
    \textbf{Advantages of Fizeau drag in DC-biased and optically-activated 3D Dirac and Weyl semimetals for nonreciprocal plasmonics}. (a) Plasmon Fizeau drag via DC bias in the 2D Dirac material, graphene. In 3D material platforms, DC bias can be applied to (b) noble metals such as Au, Al, etc., or, as we propose, to (c) 3D DSM or WSM materials. (d) The current bias shifts the Fermi disk in momentum space by $k_{shift}= k_F\mbox{sgn}[E_F] u/v_F$\cite{Sabbaghi2015Drift-inducedGraphene}, resulting in a tilted Fermi level in the illustrated electronic bandstructure\cite{Sabbaghi2015Drift-inducedGraphene,Yoon2014MeasurementGraphene}. (e) Drift-current charge carrier velocities reported for noble metals and 3D DSMs and WSMs are shown on a log scale in blue. Below, the ballistic current velocities achievable in WSMs and DSMs via the proposed PGE-enabled Fizeau drag mechanism are plotted on the same scale in red (data taken from \cite{kumar_extremely_2017, Liu2015WeylGas, behrends_visualizing_2016, Shekhar2015ExtremelyNbP, Liang2014UltrahighSemimetalCd3As2, neupane_observation_2014, Bliokh2018Electric-current-inducedPlasmon-polaritons, Kubakaddi2020DriftArsenide, Dorgan2010MobilitySiO2, Yamoah2017High-VelocityNitride}, where DSMs are labeled as squares and WSMs as diamonds). The proposed optically-activated alternative mechanism for current drag via the (f) linear PGE and (g) circular PGE in an I-symmetry breaking WSM, where ultrafast ballistic carriers are excited with either linearly or circularly polarized light, respectively. (h) Electronic bandstructure of an inversion and mirror symmetry-breaking WSM undergoing the CPGE. Nodes of opposite chirality are offset in energy. Circularly polarized light of $2|\epsilon_1| <\hbar \omega <2|\epsilon_2|$ only excites momentum-matched carriers and Pauli blocking forbids contribution from one node, creating a net photocurrent\cite{Chan2017PhotocurrentsSemimetals}. 
    }
    \label{fig:fizeau}
\end{figure*}

Finally, the application of a momentum bias instead of an external magnetic field can be used to achieve nonreciprocal optical energy transfer\cite{serra_rotating_2023, sounas_angular-momentum-biased_2014, Guo2019NonreciprocalModulation, Dong_Shen_Zou_Zhang_Fu_Guo_2015, Sohn_Kim_Bahl_2018, Kim_Kuzyk_Han_Wang_Bahl_2015}.  A practical example of a static momentum bias is plasmon Fizeau drag, where a DC current bias imparts a Doppler frequency shift on co- and counter-propagating SPP modes\cite{Borgnia2015Quasi-RelativisticGraphene,Morgado2017NegativeGraphene, DuppenCurrent-inducedGraphene, Bliokh2018Electric-current-inducedPlasmon-polaritons,Correas-Serrano2019NonreciprocalMetasurfaces,Dong2021FizeauPlasmonics,Gangaraj2022DriftingPhotonics}. Current flow breaks the Lorentz reciprocity of a material by altering the optical conductivity such that $\sigma(\mathbf{q},\omega)\neq \sigma(-\mathbf{q},\omega)$, where $\mathbf{q}$ is the wavevector and $\omega$ is the angular frequency. This approach is advantageous because it can be minimally invasive to on-chip device architectures, simply requiring a DC bias.

However, the key requirement for achieving noticeable plasmon Fizeau drag is a large carrier velocity as compared to the group velocity of the SPPs, $v_g = d\omega/dq$\cite{Dong2021FizeauPlasmonics}. For this reason, both theoretical\cite{DuppenCurrent-inducedGraphene, Borgnia2015Quasi-RelativisticGraphene, Morgado2017NegativeGraphene,Correas-Serrano2019NonreciprocalMetasurfaces, Sabbaghi2015Drift-inducedGraphene} and experimental \cite{Dong2021FizeauPlasmonics,Zhao2021EfficientGraphene} demonstrations of appreciable plasmon Fizeau drag have been limited to graphene, Fig.~\ref{fig:fizeau}(a), because it achieves exceptionally high drift carrier velocities under a DC bias\cite{Dorgan2010MobilitySiO2, Yamoah2017High-VelocityNitride, BolotinUltrahighGraphene}. All other theoretical studies consider materials with unrealistic, hypothetically relativistic charge carrier velocities\cite{Bliokh2018Electric-current-inducedPlasmon-polaritons}, when in reality carrier velocities typically saturate at much lower velocities in noble metals, Fig.~\ref{fig:fizeau}(b),\cite{Dong2021FizeauPlasmonics, Lang_2021, Bliokh2018Electric-current-inducedPlasmon-polaritons}. For example, in a copper wire with a 1 mm$^2$ cross-section under a 20 A current, charge carriers only reach a velocity of $1.6 \times 10^{-3}$ m/s\cite{Sears_1988}.

Here, we propose to use bulk WSM and Dirac semimetal (DSM) materials as a new platform to achieve appreciable plasmon Fizeau drag-induced nonreciprocity, Fig.~\ref{fig:fizeau}(c). This is motivated by the exceptionally high charge carrier mobility\cite{Gorbar2021ElectronicSemimetals,Kumar2017ExtremelyMoP2,Shekhar2015ExtremelyNbP, Kumar2021TopologicalChemistry}, Fermi velocity\cite{behrends_visualizing_2016}, and drift current density\cite{Shoron_Schumann_Goyal_Kealhofer_Stemmer_2019,Kubakaddi2020DriftArsenide, Rashidi_Shoron_Goyal_Kealhofer_Stemmer_2021} reported in some WSMs and DSMs. For example, the Fermi velocity in the DSM Cd$_3$As$_2$ has been shown to be equal to or even larger than that of graphene\cite{Liang2014UltrahighSemimetalCd3As2, Orlita2014ObservationCrystal, Timusk2013Three-dimensionalConductivity, Kubakaddi2020DriftArsenide}. Under a DC bias, the drift velocity, $v_d$, of carriers in Cd$_3$As$_2$ has been measured to reach $\sim 60\%$ of the Fermi velocity\cite{Shoron_Schumann_Goyal_Kealhofer_Stemmer_2019,Kubakaddi2020DriftArsenide, Rashidi_Shoron_Goyal_Kealhofer_Stemmer_2021}, an exceptionally high value, comparable to that seen in the graphene Fizeau drag experiments of Ref.~\cite{Dong2021FizeauPlasmonics}. These velocities are plotted in the top half of Fig.~\ref{fig:fizeau}(e), alongside those of noble metals, which are orders of magnitude smaller.

Here we present a new nonlocal optical model of plasmon Fizeau drag in 3D Dirac and Weyl materials and propose this effect as a feasible means to induce nonreciprocity. First, we present the quasi-classical description of the plasmon Fizeau drag optical response of 3D Dirac and Weyl semimetals by deriving the Fizeau drag-modified material polarizability and optical conductivity. In support of this we report the semiclassical, nonlocal derivation of the polarizability of a 3D Dirac material. We then present modeling results of plasmon Fizeau drag in the DC-biased DSM Cd$_3$As$_2$. We show that the unique blend of Fizeau drag, nonlocality, and dissipative losses in this material enables windows of pseudo-unidirectional transport, which cannot be predicted by  local models of WSM/DSM conductivities.
We show that, like in the case of magneto-optics and TRS-breaking WSMs, SPP transport can never be truly unidirectional in the face of losses and nonlocality\cite{Buddhiraju2018AbsencePlasmonics,Monticone_2020}, however, we introduce new quantitative FoMs to evaluate and rank these systems for their potential to realize pseudo-unidirectional SPP transport.

Further, we propose an alternative method to enable large Fizeau drag in WSM and DSM materials through the excitation of ultrafast carriers via the photogalvanic effect (PGE) in I-symmetry breaking WSMs and strained DSMs, Fig.~\ref{fig:fizeau}(f-h).

\section*{Results}
\subsection*{\label{sec:quasi-class-descrip} Quasi-classical description of plasmon Fizeau drag in Dirac and Weyl semimetals via drift current}

To model the effect of plasmon Fizeau drag on the optical response of bulk WSM and DSM materials, we derive a quasi-classical expression for their current-bias-modified nonlocal dynamical polarizability, $\Pi(\mathbf{q} ,\omega) \rightarrow \Pi^u(\mathbf{q},\omega)$, and the corresponding optical conductivity, $\sigma(\mathbf{q},\omega) \rightarrow \sigma^u(\mathbf{q},\omega)$, where the $u$-superscript represents the current-modified form. To the best of our knowledge, this is the first time the plasmon Fizeau drag effect in 3D WSM/DSMs is studied. As we demonstrate and discuss in detail below, properly accounting for both nonlocality and dissipative loss in 3D WSM/DSM materials is crucial for identifying the conditions for unidirectional SPP transport.  

\begin{figure} 
    \centering
    \includegraphics[width=1\linewidth]{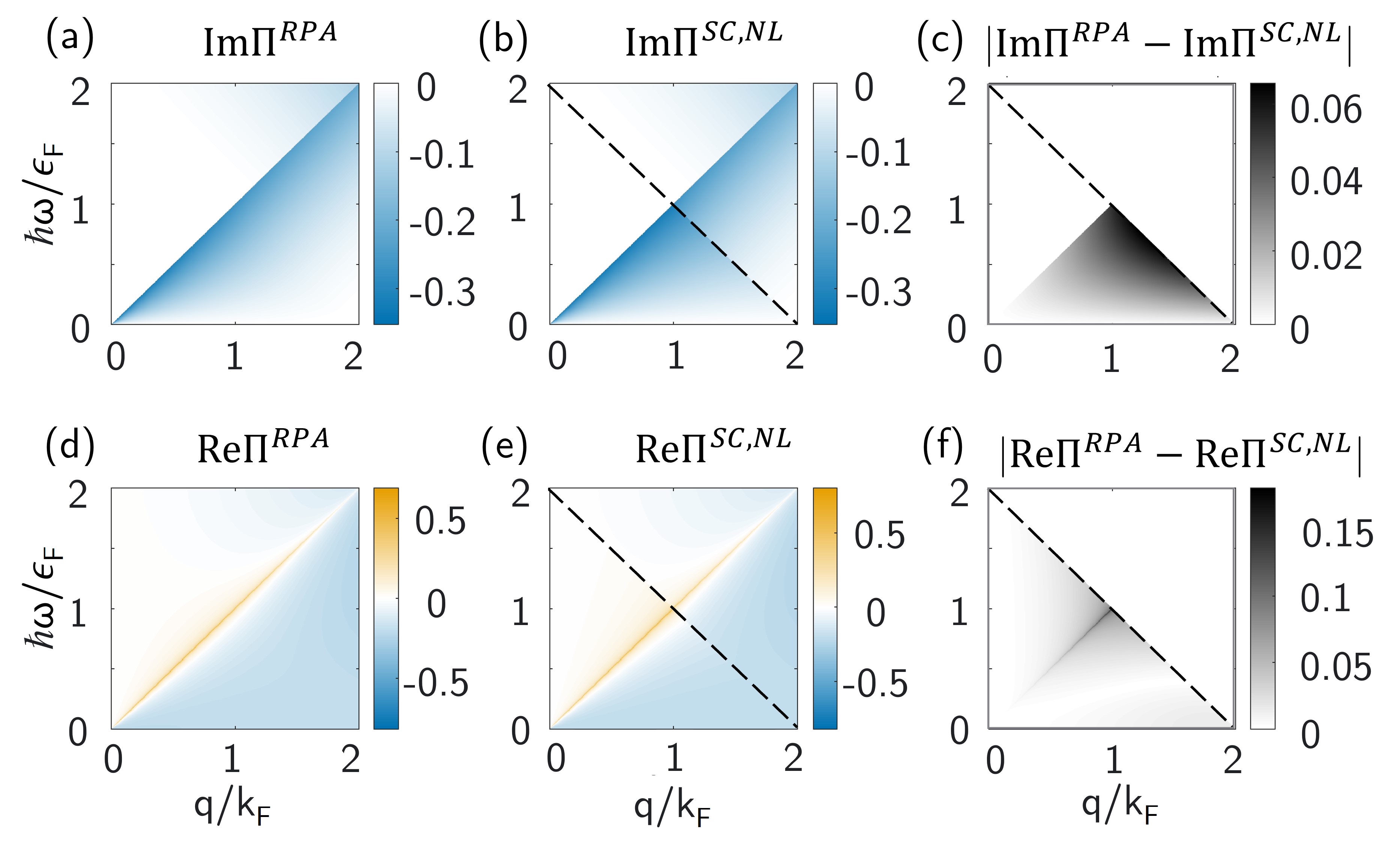}
    \caption{The longitudinal dynamical polarizability function, $\Pi(q,\omega)$, calculated via the RPA model\cite{Thakur2017DynamicalSystems, Thakur2018DynamicSemimetals} (a,d) and via the semiclassical, nonlocal (SC,NL) model introduced in this work (b,e). In (b,e), $\Pi^{SC,NL}(q,\omega)$ is plotted in the $2\mu-\hbar v_Fq-\hbar \omega>0$ domain (i.e., in the triangular area under the black dashed line), and the RPA solution is shown for comparison for $2\mu-\hbar v_Fq-\hbar \omega<0$, above the dashed line. In (c,f), the difference between the RPA and semiclassical $\Pi(q,\omega)$ is plotted. All plots are in units of $\mu^2 /\hbar^3 v_F^3$.  }
    \label{fig:pol-func}
\end{figure}

The regular, unbiased longitudinal dynamical polarizability, $\Pi(\bf{q},\omega)$ (also called the density-density response $\Pi_{\rho,\rho}(\bf{q},\omega)$, however we drop the $\rho,\rho$ subscript), and resulting dielectric function, $\varepsilon(\mathbf{q},\omega)$, of doped WSM/DSMs have been derived via the first principles linear response theory in the random phase approximation (RPA) in the literature\cite{Zhou2015PlasmonSemimetals,Thakur2017DynamicalSystems,Thakur2018DynamicSemimetals, Lv_Zhang_2013, Kotov2016DielectricFilms, Dey2022DynamicalSystem, Hofmann2015PlasmonLiquids}, reproduced in Supporting Information Note~1.

RPA is a nonlocal correlation energy functional, and thus the RPA-derived polarizability captures nonlocal optical response driven by both inter-band and intra-band electronic transitions~\cite{Mahan_2000}. However, RPA may come with a high computational cost and may suffer from convergence issues and instabilities \cite{Bates_Sensenig_Ruzsinszky_2017}. Instead, well-known, semiclassical Drude models are often used to describe longitudinal polarizability~\cite{Ziman1972PrinciplesSolids,Taylor2002SemiclassicalMetals, Bruus2004FermiTheory, JablanPlasmonicsFrequencies}. In the local approximation at zero temperature, in the absence of dissipative losses, and at frequencies below the onset of inter-band transitions, this has been shown to be\cite{Kotov2016DielectricFilms}: 

\begin{equation} \label{eq:pol-SC-local}
\begin{aligned}
    \Pi^{SC,L}({ q}, \omega) 
    & = \frac{1}{3} g N(\mu) \frac{q^2v_F^2}{\omega^2} ,   \\
\end{aligned}
\end{equation}

\noindent where $g$ is the degeneracy (for WSMs, $g=2g_W$, where $g_W$ is the number of pairs of Weyl nodes), $v_F$ is Fermi velocity, $\mu$ is the chemical potential, $N(\epsilon) = {\epsilon^2}/{2\pi^2\hbar^3v_F^3}$ is the electron density of states for a WSM/DSM, $\epsilon = \hbar \omega$ is energy, and the $SC,L$-superscript indicates semiclassical and local. Note that chemical potential is temperature dependent\cite{Coleman2015Finite-temperaturePhysics}, $\mu(\epsilon_F, T)$, and at zero temperature is equal to the Fermi energy $\epsilon_F$. 

Here, aiming to derive a computationally-efficient yet accurate approximation for the WSM/DSM polarizability and dielectric function, we follow a semiclassical framework to first derive the expressions for the unbiased $\Pi({q},\omega)$ and $\sigma({q},\omega)$ accounting for nonlocality. To the best of our knowledge, all semiclassical optical models for 3D WSM/DSMs in the literature are in the \textit{local} approximation, as in Eq.~\ref{eq:pol-SC-local}\cite{Kotov2016DielectricFilms}. As we will show, it is essential to use a \textit{nonlocal} optical model to model Fizeau drag in WSM/DSMs because transport in Dirac materials is highly nonlocal\cite{Svintsov_2018} and because the Fizeau drag induces additional, compounding nonlocality on the material\cite{Gangaraj2022DriftingPhotonics}. Thus, we provide here the first reporting of the semiclassical, nonlocal optical conductivity/permittivity for WSM/DSMs, derived via a semiclassical Boltzmann equation formalism. We find this polarizability to be

\begin{equation} \label{eq:pol-normal}
\begin{aligned}
    \Pi^{SC,NL}({{q}}, \omega) 
    &= - g N(\mu) \left(\frac{\omega}{qv_F}\frac{1}{2}\ln{\frac{\omega-qv_F}{\omega+qv_F}} + 1 \right) ,
\end{aligned}
\end{equation}

\noindent where the $SC,NL$-superscript indicates semiclassical and nonlocal. We provide the detailed derivations for $\Pi^{SC,NL}({q},\omega)$, $\sigma^{SC,NL}({q},\omega)$ and $\epsilon^{SC,NL}({q},\omega)$ in Supporting Information Note~2, and the real and imaginary parts of Eq.~\ref{eq:pol-normal} are expressed explicitly in Eq.~\ref{eq:NL-pol-Im-Re}. The $SC,NL$-superscript is dropped for conciseness in the remaining text. 
Note that the inter-band contribution can be included by adding $\sigma^{inter}(q,\omega)$ to the semiclassical conductivity\cite{Pellegrino2015HeliconsSemimetals} (Supporting Information Eq.~\ref{eq:Normal-cond-Re}), but in this work we assume low energy ranges below inter-band transitions.

In the semiclassical regime, it is assumed that the energy scales of $\epsilon = \hbar \omega$ and $v_F q$ are much less than $\mu$. To evaluate the range of applicability of Eq.~\ref{eq:pol-normal}, we compare its prediction of the polarizability function to the corresponding RPA result over the $2\mu-\hbar v_Fq-\hbar \omega>0$ domain in Fig.~\ref{fig:pol-func}. The comparison of the two models shows their reasonable agreement. As an additional validation of the new model, the unbiased SPP dispersion is plotted for the Cd$_3$As$_2$ case (Supporting Information Fig.~\ref{fig:NL-L_RPA-SPP-comparisons}), also matching well with the results of the RPA model. It should be noted that the RPA and semiclassical expressions for the polarizability (i.e., Eqs.~\ref{eq:pol-RPA}, \ref{eq:pol-SC-local}, and \ref{eq:pol-normal}) are valid in the low-temperature regime when $kT\ll \epsilon_F$. 

We then derive the current-biased expressions $\Pi^u({q},\omega)$ and $\sigma^u({q},\omega)$, which account for a current-modified charge density distribution in the material and enable estimations of the Fizeau drag effect on the SPP modes propagation. This derivation is outlined in detail in Supporting Information Note~2. Similarly to the unbiased case, we make use of Boltzmann transport and linear response theory, however, the equilibrium Fermi sphere is now shifted in momentum space and the Fermi distribution, as shown in Fig.~\ref{fig:fizeau}(d), and becomes\cite{Dong2021FizeauPlasmonics, Borgnia2015Quasi-RelativisticGraphene}:

\begin{equation} \label{eq:dist}
    f^{u}(\epsilon) = \frac{1}{e^{(\epsilon(\mathbf{p})-{\mathbf{u\cdot p}}- \mu^u )/k_BT}}.
\end{equation}

\noindent Here, $\mathbf{u}$ is the drift velocity of the quasiparticles, $\mathbf{p}$ is the quasiparticle momentum, $\epsilon(\mathbf{p})$ is energy, and $\mu^u$ is a modified chemical potential. The effect of Eq.~\ref{eq:dist} on the material polarizability and optical conductivity is then propagated through the derivation. 

We show that the electromagnetic and electron frequency and momentum in the WSM/DSM longitudinal polarizability are transformed under a quasi-Lorentzian transform (QLT) when Eq.~\ref{eq:dist} is applied in the $\Pi^u({q},\omega)$ derivation.

Due to the linearly dispersing electron bands, $\epsilon({\bf p}) = \mbox{sgn}(\mu) v_F|{\bf p}|$, massless Weyl and Dirac dispersions are characterized by a \textit{quasi-Lorentz invariance}, where the Fermi velocity $v_F$ replaces the speed of light in the usual Lorentz factor, $\gamma = (1 - u^2/v_F^2)^{-1/2}$, as has been discussed for the Dirac dispersion of graphene \cite{Borgnia2015Quasi-RelativisticGraphene, Svintsov2013HydrodynamicGraphene, Shytov_2009, Kotov_Uchoa_Pereira_Guinea_Castro_Neto_2012}. 
The form of $\epsilon({\bf p}) = \mbox{sgn}(\mu) v_F|{\bf p}|$, for ${\bf u} \parallel {\bf q} \parallel \mathbf{\hat{z}}$ is preserved under the QLT\cite{Peskin1995AnTheory.}

\begin{equation}
\begin{aligned}
  \epsilon_0(p)= \gamma \left( \epsilon(p)-up^z \right)&,~
    p_0^z=\gamma\left(p^z-\frac{u}{v_F^2}\epsilon(p)\right)&, \\
    p_0^y=p^y&,~p_0^x=p^x,
\end{aligned}
\end{equation}
where the $0-$subscript designates the moving frame.

This is in contrast to materials with parabolic dispersions, which have Galilean-invariant, nonrelativistic dispersion relation $E=p^2/2m$ \cite{Abedinpour2011DrudeSheets}. 
The QLTs of the electromagnetic parameters are 
\begin{equation}
\begin{aligned} 
        \omega_0=\gamma(\omega -u q^z)&,~
        q_0^z=\gamma(q^z-\frac{u}{v_F^2}\omega), \\
        q_0^y=q^y&,~q_0^x=q^x.
\end{aligned}
\end{equation}
 The perturbed carrier density can be rewritten as

\begin{equation} \label{eq:QLT-Fermi-dist}
    f^{u}(\epsilon) = \frac{1}{e^{(\epsilon-u p^z- \mu^u )/k_BT}+1} = \frac{1}{e^{(\epsilon_0 -\mu_0)/\gamma k_BT}+1} ,
\end{equation}
where we have introduced the transformed chemical potential $\mu_0 =\gamma \mu^u$. To conserve carrier density the chemical potential must transform as $\mu_0 = \mu/\gamma^{1/3}$ (Supporting Information Eq.~\ref{eq:QLT-density}). 
By substituting the QLT for the electromagnetic and electric quantities due to current traveling in the z-direction into the derivation for $\Pi^u({q},\omega)$, we see that this ansatz for the transformation greatly simplifies the derivation and reveals the QLT nature of current bias in DSM/WSMs.
The longitudinal polarizability in a current carrying state is found to be:
\begin{equation} \label{eq:pol-Fizeau}
\begin{aligned}
    \Pi^{u}({q}, \omega) 
    & = - g N(\mu_0) 
     \frac{q^2}{q_0^2} 
     \left( \frac{\omega_0}{2 q_0 v_F} \ln \left( \frac{\omega_0 - q_0 v_F}{\omega_0 + q_0 v_F}\right) +1
    \right) .
\end{aligned}
\end{equation}

\noindent Transforming Eq.~\ref{eq:pol-Fizeau} according to ${\sigma^u} ({q},\omega) = (ie^2\omega/q^2) \Pi^u ({q}, \omega)~$\cite{Bruus_Flensberg_2004}, the current-biased, longitudinal optical conductivity is transformed via the relation

\begin{equation} \label{eq:Fizeau-transform}
\begin{aligned} 
        \sigma^{u} (\omega,{q},\mu)
        &=  \frac{\omega}{\omega_0}\sigma_{zz}(\omega_0,{q_0},\mu_0) . \\
\end{aligned}
\end{equation}

\noindent 
From $\sigma^u$, the impact of DC bias Fizeau drag on SPPs at the surfaces of 3D WSM/DSM materials can be modeled. Note that because we are considering TRS-preserving WSMs for Fizeau drag, we do not solve for or consider the Fizeau drag impact on the anomalous Hall effect or $\sigma_{AHE}$ in this work.

\subsection*{Photogalvanic Effect enabled plasmon Fizeau Drag} \label{sec:CPGE}

While the exceptional mobility and current density of some 3D WSM/DSMs already make them good candidates for Fizeau drag via DC bias, we propose an additional Fizeau drag mechanism via the PGE, Fig.~\ref{fig:fizeau}(f-g). The linear PGE (LPGE), also called shift current, is manifested as the shift in space of a charge carrier during the inter-band photoexcitation process under illumination by linearly polarized light, stemming from the I-symmetry breaking of the material\cite{Dai_Rappe_2023}.  Berry flux fields contribute to shift current, making WSMs attractive material candidates for observing this effect\cite{Liu2020SemimetalsPhotodetection, Weng_2019, Osterhoudt_Colossal}. In the circular PGE (CPGE), illumination from circularly polarized light selectively excites charge carriers with preferred linear momenta, Fig.~\ref{fig:fizeau}(h), supporting a net photocurrent in certain symmetry groups of I-symmetry-breaking WSM or strained DSM materials\cite{DeJuan2017QuantizedSemimetals, Rees2020Helicity-dependentRhSi,Liang2022Strain-inducedSubstrate}. Importantly, the excited ballistic carriers propagate with velocity $v_F$ over their mean free path\cite{Dai_Rappe_2023}. 

By taking advantage of the ultrafast ballistic carriers continuously generated via the PGE mechanism, we can bypass the carrier velocity limits achievable with drift current via DC bias. We propose that the PGE photoexcited carriers can impart Fizeau drag on the electron plasma, thus providing an external mechanism to control nonreciprocal behavior by optical pumping. The $v_F$ is inherently high in many WSM and DSM materials, while the saturation velocity in the DC-biased regime is limited by material synthesis, Fig~\ref{fig:fizeau}(e). PGE mechanism could thus allow observations of strong Fizeau drag in many more material candidates than what is achievable via DC bias. 

As shown in Fig.~\ref{fig:fizeau}(g), the CPGE current mechanism is inherently different from DC bias, Fig.~\ref{fig:fizeau}(d), in the electron bandstructure picture, resulting in different equilibrium charge distributions. Thus, the optical model derived using Eq.~\ref{eq:dist} is inherently only applicable to the DC-biased case; a different treatment of the equilibrium carrier distribution should be used for the PGE case which accounts for the relative excited band populations. However, related THz-pumping experiments in topological insulators (TIs) have shown that the DC bias modeling done here may, in fact, provide a good approximation for the PGE-induced Fizeau drag when modified to consider pumping dynamics.

In THz pumping experiments of the TI Bi$_2$Te$_3$, ultrafast and strong fields were found to accelerate the Dirac surface states, in a phenomenon called light-field-driven currents\cite{Reimann_Schlauderer_et}, which is distinct from, but related to the PGE\cite{Kiemle_Zimmermann_Holleitner_Kastl_2020}. Under a semiclassical approximation, the current density was $j\approx-e v_f n \Delta k(t)/ k_F$, where $\Delta k(t)$ was the  shift in the electron density observed in the direction of the current via angle-resolved photoemission spectroscopy (ARPES). The time-dependent electron distribution of the Dirac cone was approximated, assuming a small relaxation time $\tau_R$, as 
\begin{equation} \label{eq:THz-pump-dist}
    \begin{aligned}
        f(\epsilon,t)
        &\approx
        \frac{1}{e^{(\epsilon(\mathbf{p}) - \mathbf{v_F} \cdot \Delta \mathbf{p}(t) - \mu)/k_BT}+1},
    \end{aligned}
\end{equation}
where $\Delta \mathbf{p}(t) = \hbar \Delta \mathbf{k}(t)$,  is a function of  $\tau_R$ and the pumping field $E(t)$. This semiclassical description matched the experimental results for Bi$_2$Te$_3$ THz pumping experiments, and $\Delta \mathbf{k}(t)$ was directly proportional to the temporal current density measurement. Compared to Eq.~\ref{eq:dist}, we see that Eq.~\ref{eq:THz-pump-dist} is equivalent with the exception that the $\Delta \mathbf{k}$ terms have different physical origins and that the latter is time dependent. 

Both the DC-biased, Eq.~\ref{eq:dist}, and PGE, Eq.~\ref{eq:THz-pump-dist}, cases can be considered as cases where the chemical potential, $\mu$, becomes band-dependent
\begin{equation} \label{eq:band-dep-mu}
    \begin{aligned}
        \mu(\epsilon_F, T) &\rightarrow \mu(\epsilon_F, T, \delta n) . \\
    \end{aligned}
\end{equation}
Under DC biasing, $\mu(\epsilon_F, T, \delta n) \approx \mu^u(\epsilon_F, T)  + \mathbf{u} \cdot \mathbf{p}$ (Eq.~\ref{eq:QLT-Fermi-dist}), and under photoexcitation $\mu(\epsilon_F, T, \delta n) \approx \mu(\epsilon_F, T)  + \mathbf{v}_F \cdot \Delta \mathbf{p}(t)$. In both cases, the electron density is modified by an external perturbation $\delta n$: in the case of drift current $\delta n$ is due to a voltage drop across the material, while in the case of optical pumping, $\delta n$ is a result of photoexcitation\cite{Tomadin_Polini_2021}. 

Similar ARPES experiments could be conducted for the transient electron distribution in WSM/DSMs under the PGE. From these experiments, a similar semiclassical treatment of the PGE pumping in WSM could be performed. Such results for $\Delta \mathbf{k}(t)$ of a 3D Dirac or Weyl material under the PGE could extend the analysis performed here for the DC drift-biased case to the PGE pumping case.

It is important to acknowledge that - unlike the DC-biased case - the PGE Fizeau drag approach requires an external pumping field, but in principle can be activated without the electrodes, which simplifies the on-chip design and fabrication requirements. However, under prolonged pumping, a generated space-charge field can lead to so-called 'optical damage’\cite{Tan_Zheng_Young_Wang_Liu_Rappe_2016}, which can be alleviated by adding the electrodes similar to the DC-biased case. 
Overall, the proposed alternative way to achieve the Fizeau drag holds interest in studying carrier dynamics and interactions in a wide range of Weyl and Dirac materials and can be used to develop new techniques to achieve and models to describe the alteration of symmetry at ultrafast timescales. This mechanism could add to the toolbox of THz-driven controls of quantum materials and studies of many-body dynamics\cite{Yang_Li_Fiebig_Pal_2023}.

\subsection*{Fizeau-drag SPP dispersion in 3D Weyl and Dirac semimetals} \label{sec:Cd3As2-results}

We model plasmon Fizeau drag under drift current bias using the transformed nonlocal Drude model in Eq.~\ref{eq:pol-Fizeau}. We look at the example case of Cd$_3$As$_2$, a 3D DSM, which has ultrahigh experimentally-measured carrier mobility, $\mu=9\times10^6$ cm$^2$V$^{-1}$s$^{-1}$and Fermi velocity, $v_F=9.3-1.5\times10^5$ m s$^{-1}$\cite{Liang2014UltrahighSemimetalCd3As2}. The current density has been reported to exceed $5~\mbox{A/mm}$ for a carrier density of $5\times10^{12} \mbox{cm}^{-2}$, corresponding to an exceptionally high drift velocity of $u=6.3\times10^5$ ms$^{-1}$ $\approx 0.6 v_F$\cite{Shoron_Schumann_Goyal_Kealhofer_Stemmer_2019}. Epitaxially grown thin films of Cd$_3$As$_2$ have also demonstrated high mobility and $v_F$\cite{Uchida_Nakazawa_Nishihaya_Akiba_Kriener_Kozuka_Miyake_Taguchi_Tokunaga_Nagaosa_2017, Nakazawa_Uchida_Nishihaya_Kriener_Kozuka_Taguchi_Kawasaki_2018, Rice_Nelson_Norman_Walker_Alberi_2022, Rice_Lee_Fluegel_Norman_Nelson_Jiang_Steger_McGott_Walker_Alberi_2022}.

\begin{figure*} 
    \centering
    \includegraphics[width=12.9cm]{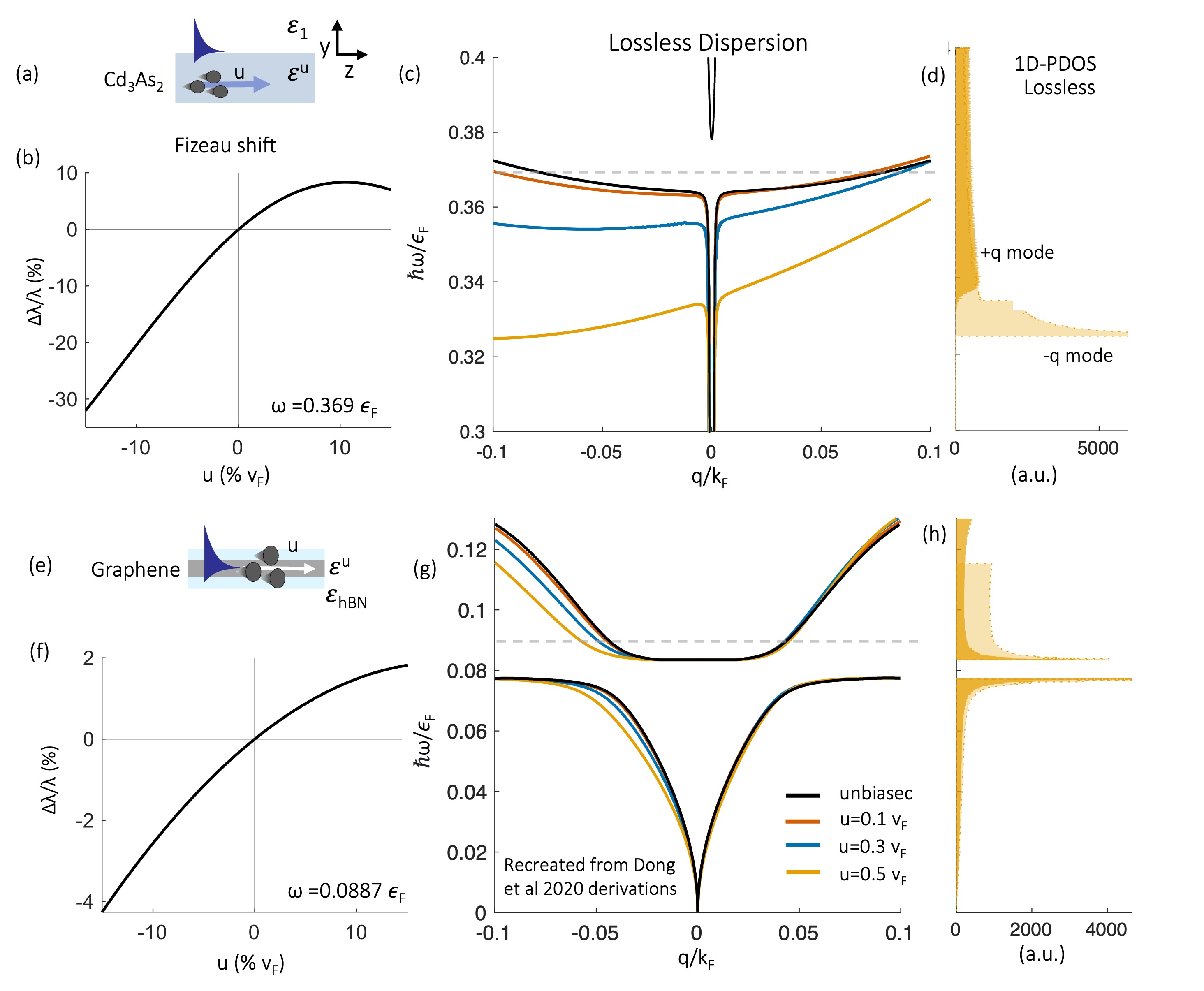}
    \caption{ \textbf{Plasmon Fizeau drag in Cd$_3$As$_2$ and graphene under the lossless approximation}
    (a) The nonreciprocity of the lossless SPP modes under Fizeau drag at the interface between bulk DSM Cd$_3$As$_2$ material and air is represented as (b) Fizeau shift $=(\Delta\lambda)/\lambda_0$ plotted as a function of bias drift velocity at frequency $\omega=0.369\epsilon_F$.
    (c) The SPP dispersion, which is calculated in the lossless regime with and without drift bias ($\epsilon_F = 0.15$ eV, $v_F=1\times10^6$ ms$^{-1}$, $\varepsilon_{\infty}=13$, $\varepsilon_b = 12$, $g=4$\cite{Kotov2016DielectricFilms}). The upward bending of the SPP modes at high q values is a characteristic signature of nonlocality. The frequency $\omega=0.369\epsilon_F$ at which the Fizeau shift values were calculated in (b) corresponds to the dashed line in (c). 
    (d) Alternatively, nonreciprocity and unidirectionality of SPPs can be represented using the calculated 1D-PDOS values of SPP modes, which reveal that the counter-propagating mode dominates at lower energies as the drift velocity increases. 
    Note that we truncate our 1D-PDOS calculation at the momentum onset of Landau damping. (e-h) The Fizeau shift (f), SPP dispersion (g) and 1D-PDOS (h) for Fizeau drag in Graphene encapsulated in hBN\cite{Lin2017All-angleHeterostructures, Dong2021FizeauPlasmonics} is shown for comparison. Graphene is modeled following the equations in Ref.~\cite{Dong2021FizeauPlasmonics}, using $n=2.88\times 10^{12}$, $v_F=1\times 10^6$ms$^{-1}$, $E_F=198$ meV. The hBN is modeled using the parameters from Ref.~\cite{Ni_McLeod_Sun_Wang_Xiong_Post_Sunku_Jiang_Hone_Dean_2018}.}
    \label{fig:fizeauResults}
\end{figure*}

We model the system assuming that the drift current is in the longitudinal direction, corresponding to the z-axis as illustrated in Fig.~\ref{fig:fizeauResults}(a). The longitudinal optical conductivity, $\sigma_L = \sigma_{zz}$, is transformed according to the QLT that we derived in Eq.~\ref{eq:Fizeau-transform}. The transverse optical conductivity,  $\sigma_T=\sigma_{yy}$, is expected to exhibit a weaker transformation, which is neglected here for simplicity, $\sigma_T^u \approx \sigma_T$ \cite{Borgnia2015Quasi-RelativisticGraphene,Morgado2022DirectionalGraphene, Gangaraj2022DriftingPhotonics}. SPP eigenmodes corresponding to the SPPs propagating along the interface of the anisotropic DSM and air are computed by using the surface boundary conditions obtained from Maxwell’s equations. Assuming p-polarized surface modes of the form $E_1 = [0~E_{1,y}~E_{1,z}] e^{(iqz-\kappa_1y)}$ in the top layer,  $E_2 = [0~E_{2,y}~E_{2,z}] e^{(iqz+\kappa_2y)}$ in the DSM and continuity of $E_{||}$ and $H_{||}$ at y=0, the resulting SPP dispersion relation is\cite{Gangaraj2022DriftingPhotonics}:

\begin{equation}
    q = \pm \frac{\omega}{c} \sqrt{ \varepsilon_1 \frac{\varepsilon^u_T(\varepsilon_L^u-\varepsilon_1)}{\varepsilon_L^u \varepsilon_T^u - \varepsilon_1^2} }.
    \label{eq:dispersion}
\end{equation}
Here the permittivity is related to the optical conductivity in the intra-band model according to $\varepsilon_{ij}(q,\omega)=\varepsilon_{\infty}(\omega) + i  {\sigma_{ij}(q,\omega)}/{\varepsilon_0\omega} $ \cite{Chen2019OpticalSemimetals, Kotov2016DielectricFilms}  and $\varepsilon_1$ describes the permittivity of the medium above the WSM/DSM, which in this case is air, $\varepsilon_1=1$. To numerically solve for the SPP modes, we search for solutions across a complex $q=\mbox{Re}[q]+i\mbox{Im}[q]$ plane, as described in Supporting Information Note~4, using a root-finding protocol such that the real and imaginary parts of $q$ of the modes are solved for.

\subsubsection*{Comparison to graphene in the lossless approximation}

First, we consider the lossless case for carriers with drift velocity, $v_d=u$, varying between $10\%$ and $50\%$ of $v_F$, Fig.~\ref{fig:fizeauResults}(a). 
With increased drift velocity, the dispersion branch corresponding to the counter-propagating surface mode (i.e., the mode with the $-q$-wavevector direction in this configuration) is progressively red-shifted. Such Fizeau-drag-induced Doppler red-shift has also been predicted for SPPs on metals\cite{Gangaraj2022DriftingPhotonics} and demonstrated experimentally for graphene SPPs\cite{Dong2021FizeauPlasmonics, Zhao2021EfficientGraphene}.
Additionally, an inflection point appears in the counter-propagating mode dispersion curve, at which point the group velocity of the SPP is reversed\cite{Gangaraj2022DriftingPhotonics}, similar to the case of noble metals. This inflection point is however not observed for the DC-biased SPP modes in graphene, Fig.~\ref{fig:fizeauResults}(g)\cite{Dong2021FizeauPlasmonics}.  

For $q$ values much larger than the shift of the Fermi disk, $k_{shift}= k_F  \mbox{sgn}[E_F] u/v_F$\cite{ Sabbaghi2015Drift-inducedGraphene}, the SPP dispersion must converge with the normal, un-biased SPP mode dispersion, plotted in black in Fig.~\ref{fig:fizeauResults}(a). In Figs.~\ref{fig:fizeauResults}(a) and (d), $q=k_{shift}$ is at the very edge of the x-axis for the case of $u=0.1v_F$ ($q= k_{shift}=0.1 v_F$) and outside the bounds of the x-axis for the $u=0.3,0.5 v_F$ cases, so that such converging behavior is not visible. In an expanded q-axis plot, clear convergence of the dispersion plots at $q > k_{shift}$ is observed, Fig.~\ref{fig:q-to-kshift}. 

For direct comparison between our findings in the 3D WSM/DSM case and past results in graphene (Fig.~\ref{fig:fizeauResults}(e)), we also plot the dispersion characteristics of SPP modes in Fizeau drag biased graphene encapsulated in hBN in Figs.~\ref{fig:fizeauResults}(f-h), which have been calculated using the derivations within Ref.~\cite{Dong2021FizeauPlasmonics}.

To quantify nonreciprocity of the SPP mode in the lossless case,  past works have used the Fizeau shift, the change in the SPP wavelength with current bias ($(\Delta\lambda)/\lambda_0$), as the figure of merit (FoM)\cite{Zhao2021EfficientGraphene,Dong2021FizeauPlasmonics}, where $\lambda_0$ is the SPP wavelength without bias. 
This FoM is plotted in Figs.~\ref{fig:fizeauResults}(b) and (f) for the Cd$_3$As$_2$ and graphene cases, respectively, as a function of the DC bias drift carrier velocity\cite{Dong2021FizeauPlasmonics}. We evaluate the Fizeau shift in Cd$_3$As$_2$ at $\omega =0.369 \epsilon_F = 13.4$~THz, which corresponds to unbiased $\lambda_0=360 $ nm in Fig.~\ref{fig:fizeauResults}(c). These results are qualitatively comparable to the Fizeau shift measured in graphene, Fig.~\ref{fig:fizeauResults}(f), and reveal an almost order-of-magnitude improvement in the FoM value relative to graphene at the same drift velocities\cite{Dong2021FizeauPlasmonics} for $\omega=0.0887 \epsilon_F$. However, comparing the dispersion plots for the 3D DSM and graphene, we can conclude that this FoM does not adequately estimate the effect of the nonreciprocity on achieving unidirectional SPP transport, and thus does not allow for proper comparison between different material systems and external biases by this metric. 

While both dispersion curves in Fig.~\ref{fig:fizeauResults} and exhibit significant nonreciprocity, the capability of 2D and 3D Dirac materials to achieve a unidirectional SPP transport is hindered by nonlocality in different ways. In both cases, true unidirectional transport is not possible due to nonlocality, and the counter-propagating mode experiences red shift. However, differently from the 2D graphene case, the inflection point observed in the 3D DSM $-q$ SPP branch opens up a narrow window of a pseudo-unidirectional SPP transport in the counter-propagation direction, evident for $u=0.3v_F$ and $u=0.5v_F$ in Fig.~\ref{fig:fizeauResults}(c).

To properly evaluate the nonreciprocity and its effect on achieving highly-asymmetrical SPP transport within some frequency windows, we propose to use a different FoM by calculating and comparing the density of photon states available for the SPP modes co-propagating and counter-propagating with the bias current. This universal FoM - The 1D photonic density of states (1D-PDOS) - is calculated by counting the available states in the momentum (q-vector) space corresponding to a narrow frequency window sliding along the frequency axis:

\begin{equation}
    \mbox{1D-PDOS}(\omega)= \int_{\omega - \delta \omega}^{\omega+\delta \omega} q(\omega) d \omega.
\end{equation}

1D-PDOS is calculated numerically by separately integrating over the dispersion curves in the forward and reverse directions. 
Note that for this comparison we integrate 1D-PDOS in the momentum space up to the onset of Landau damping.

The results for Cd$_3$As$_2$ Fizeau drag with $u=0.5v_F$ bias velocity plotted in Fig.~\ref{fig:fizeauResults}(d) show that, in this lossless case, a low energy frequency window can be opened where the counter-propagating mode (light yellow) exhibits a much higher 1D-PDOS than the co-propagating one (dark yellow), effectively allowing a pseudo-unidirectional SPP transport. 
This circumstance differs notably from the prediction of truly unidirectional SPP transport through the co-propagating SPP mode, which can be made when invoking the non-realistic local Drude mode (see Supporting Information Fig.~\ref{fig:NL-L_RPA-SPP-comparisons}). 
In the nonlocal case shown in Fig.~\ref{fig:fizeauResults}(c-d), both, co- and counter-propagating modes are allowed within the pseudo-unidirectional transport window, but the amount of energy that can be back-scattered to the co-propagating one is severely limited by its very low DOS.

Further, the 1D-PDOS shows that the graphene and DSM/WSM cases have qualitatively very different nonreciprocal response under plasmon Fizeau drag. The lossless graphene case has very little contrast in 1D-PDOS between the co- and counter-propagating SPP modes. Thus, windows of pseudo-unidirectional transport are not observed in graphene in the lossless case, but are predicted in DSM/WSMs.

Finally, to get the full physical picture of nonreciprocal propagation of SPP modes under current bias in 3D DSM and WSM materials, dissipative losses must be included in the analysis together with the nonlocality.

\subsubsection*{Combined effect of dissipation and nonlocality on nonreciprocal SPP transport}

\begin{figure*} 
    \centering
    \includegraphics[width=16cm]{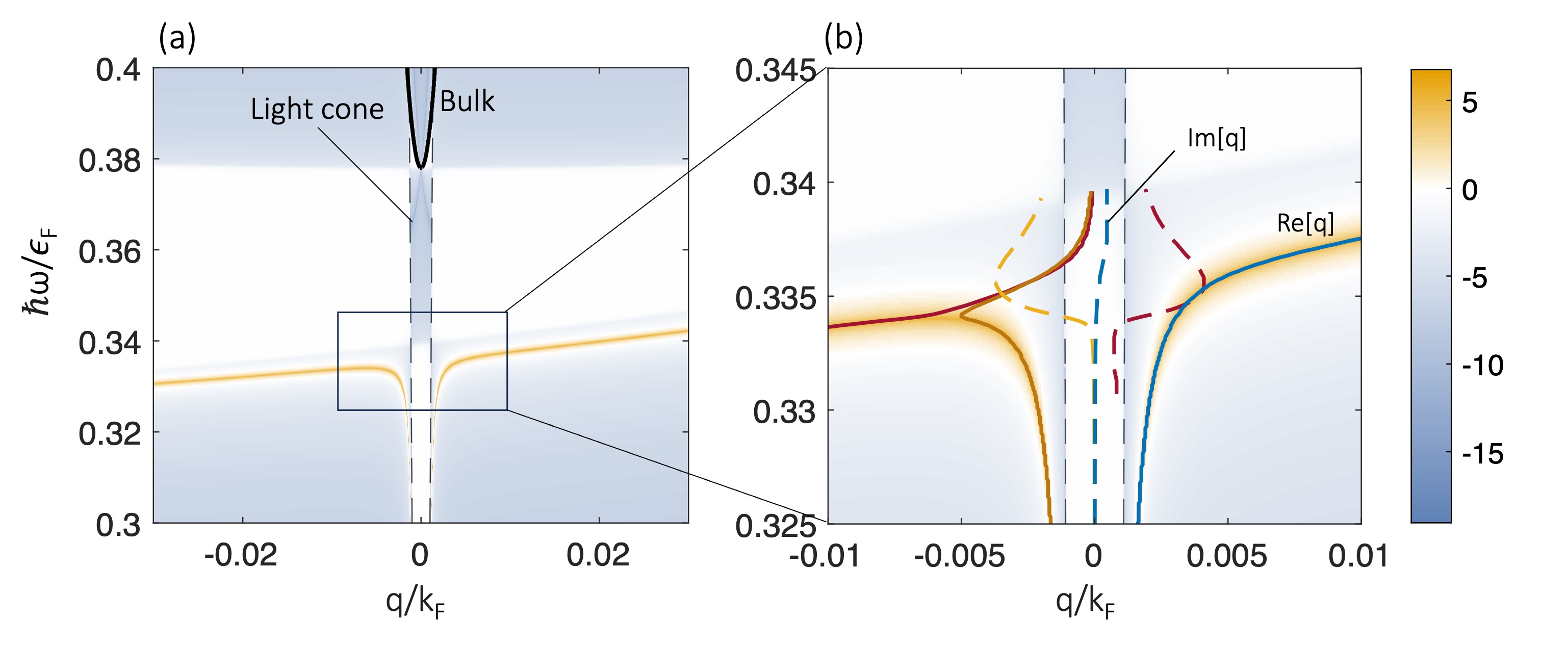}
    \caption{ \textbf{Loss Function and SPP dispersion accounting for relaxation.} 
    (a) Loss function, $L(q,\omega)=\mbox{Im}~r_{pp}$, of the semi-infinite Cd$_3$As$_2$-air interface under current bias of drift velocity $u=0.3v_F$ with $\tau=2.1\times10^{-11}$ s plotted in log units. (b) The dispersion curves of the SPP modes found as the solutions of the complex dispersion relation Eq.~\ref{eq:dispersion} are overlaid on $\mbox{Im}~r_{pp}$. The solid lines and dashed lines represent the $\mbox{Re}[q]$ and $\mbox{Im}[q]$ part of each complex solution, respectively. Note that one of the $-q$ propagating SPP modes (red) is characterized by $\mbox{Im}[q] \cdot \mbox{Re}[q]<1$, corresponding to a gain mode, which is discussed in Supporting Information Note~4 alongside discussions of the continuum mode\cite{Echarri_Gonçalves_Tserkezis_Abajo_Mortensen_Cox_2021,Costa_Gonçalves_Basov_Koppens_Mortensen_Peres_2021,Park2022PlasmonicNodes} represented as the white band above the SPP mode. For the remainder of the manuscript, we neglect this gain mode as unphysical, as discussed for the case of current-biased SPP modes on noble metal interfaces in Ref.~\cite{Monticone_2020}. The stipes within the light cone represent a change in the sign of $\mbox{Log}(\mbox{Im}~r_{pp})$. The same parameters for Cd$_3$As$_2$ are used as in Fig.~\ref{fig:fizeauResults}.
    }
    \label{fig:lossyFunc}
\end{figure*}

\begin{figure*} 
    \centering
    \includegraphics[width=12.9cm]{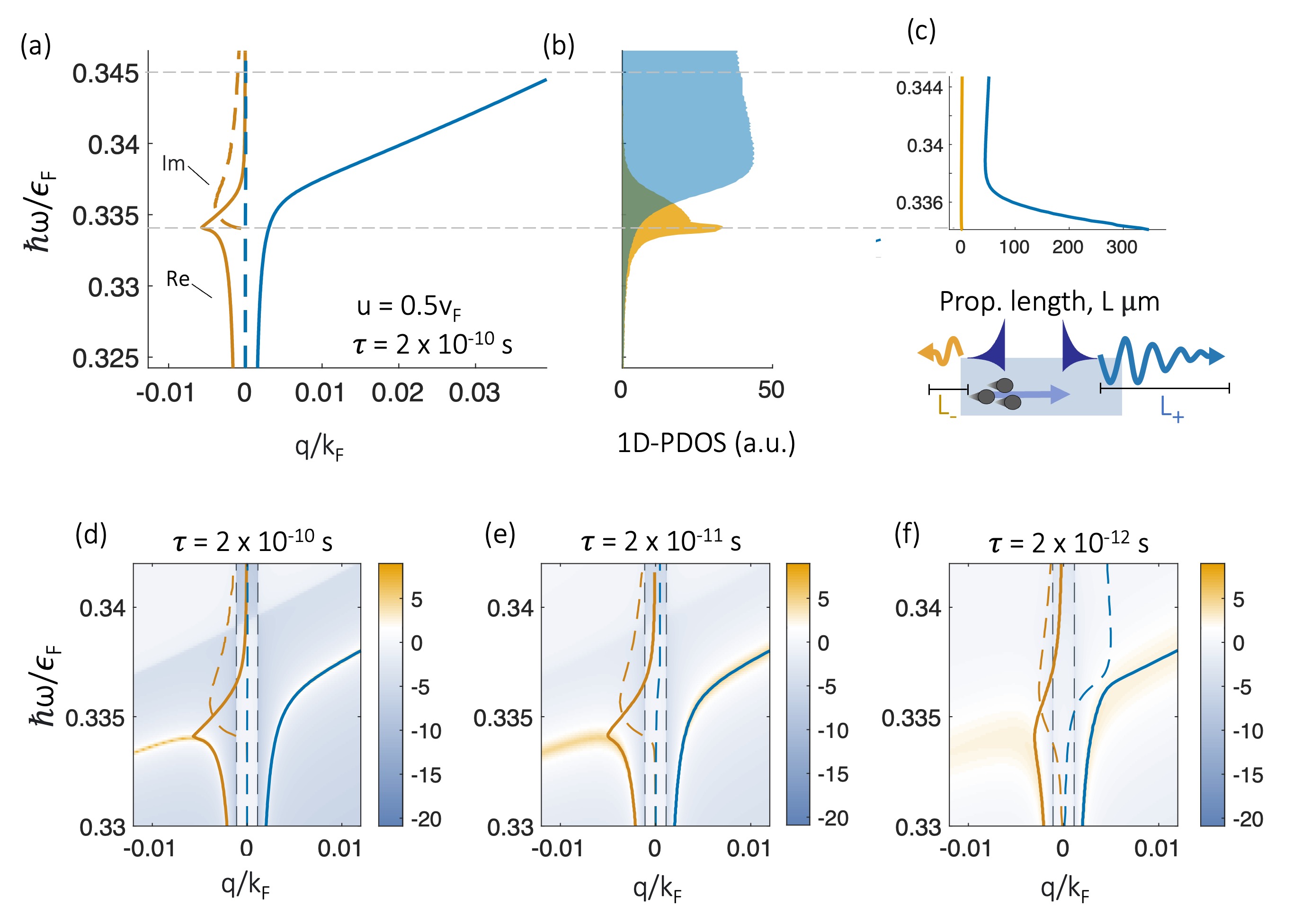}
    \caption{ \textbf{Impact of finite damping on nonlocal SPP modes under a DC bias.} (a) For a scattering period of $\tau = 2.1\times 10^{-10}$s and a DC bias of  $u=0.5v_F$, loss causes band-bending of the counter-propagating, $-q$-wavevector mode (yellow) yet does not impact the co-propagating, $+q$-wavevector mode (blue). The counter-propagating mode is also severely damped, as evident from the much larger magnitude of its imaginary component of the wavevector (compare blue and yellow dashed lines). (b) The corresponding 1D-PDOS values and (c) propagation lengths for the co- and counter-propagating modes show that the co-propagating mode dominates in both 1D-PDOS and propagation length at energies above 0.338$\epsilon_F$. In (d-f) the evolution of the SPP dispersion is illustrated as $\tau$ is progressively decreased, with the complementary $\mbox{Re}[q]$ and $\mbox{Im}[q]$ components plotted as solid and dashed lines, respectively. A heatmap of the loss function $\mbox{Im}~r_{pp}$ is plotted in log units. In (d-f) the co-propagating mode is preserved and underdamped, while the counter-propagating mode experiences increased back-bending and overdamping. The same parameters for Cd$_3$As$_2$ are used as in Fig.~\ref{fig:fizeauResults}.
    }
    \label{fig:lossy-result}
\end{figure*}

Dissipation is an inherent property of any plasmonic media\cite{Boriskina2013PlasmonicApplications} and the impacts of finite damping on the system must be considered. Dissipative losses cause folding of the SPP dispersion curves away from their asymptotic behavior and thus erase the nonreciprocal modes at large wavevectors. However, in some cases, nonlocality can open frequency windows where the SPP transport remains bidirectional but one of the modes is significantly more overdamped than the other\cite{ Gangaraj2022DriftingPhotonics}.
Cd$_3$As$_2$ is an excellent candidate to explore this effect due to its exceptionally high electron transport lifetime: $\tau = 2.1 \times 10^{-10}$s\cite{Liang2014UltrahighSemimetalCd3As2}. 

In Fig.~\ref{fig:lossyFunc}(a-b) we solve for the loss function, $L(q,\omega)=\mbox{Im}~r_{pp}$\cite{Costa_Gonçalves_Basov_Koppens_Mortensen_Peres_2021}, of the semi-infinite Cd$_3$As$_2$-air interface under a current bias of drift velocity $u=0.3v_F$ with $\tau=2.1\times10^{-11}$s and overlay the complex SPP mode solutions, where $r_{pp}$ is the p-polarized reflection coefficient\cite{Pajovic_2021}. The loss function peaks indicate underdamped modes that can be easily excited. Here, the co-propagating mode (blue) is underdamped as evidenced by the loss function and the complex mode solution which does not exhibit back-bending and follows $|\mbox{Re}[q]|>|\mbox{Im}[q]|$. However, the counter-propagating mode (yellow) does experience loss-induced back-bending behavior and is overdamped as evidenced by $|\mbox{Im}[q]|>|\mbox{Re}[q]|$ in the complex mode solution. Note that another $-q$ propagating SPP modes (red) is present in the complex solution but is characterized by $\mbox{Im}[q] \cdot \mbox{Re}[q]<1$, corresponding to a gain mode. This gain mode is discussed in Supporting Information Note~4 and we neglect it as unphysical for the remainder of the manuscript,  as discussed for the case of Fizeau drag at noble metal interfaces in Ref.~\cite{Gangaraj2022DriftingPhotonics}.

In Fig.~\ref{fig:lossy-result}(a) we show the complex solutions of the dispersion equation for the high drift velocity and high mobility case of $u = 0.5v_F$ and $\tau = 2.1 \times 10^{-10}$s, parameters corresponding to the best experimentally-confirmed material properties of Cd$_3$As$_2$. This shows the same imbalance of an overdamped counter-propagating mode (yellow) and underdamped co-propagating mode (blue) with a more pronounced group velocity than in Fig.~\ref{fig:lossyFunc}(b).
The 1D-PDOS in this situation, shown in Fig.~\ref{fig:lossy-result}(b), is very different from the lossless case in Fig.~\ref{fig:fizeauResults}(d). Here, the counter-propagating mode again dominates at lower energies, but now the overdamping of that mode opens a higher frequency window where the co-propagating mode dominates.

In addition to the 1D-PDOS, we also evaluate and compare the propagation lengths of both SPP modes. The propagation length is defined here as the distance over which the co- or counter-propagating SPP mode power decays by a factor of $1/e$, $L_{\pm}=1/(2\mbox{Im}[q])$\cite{Berini_2009}. Fig.~\ref{fig:lossy-result}(c) shows that, in the frequency window designated by the gray dashed lines, the counter-propagating mode has almost zero propagation length as compared to the co-propagating mode, whose propagation length can span from tens to hundreds of microns.
At lower frequencies, however, where the counter-propagating mode exhibits a much higher 1D-PDOS, a high level of propagation loss would be expected if the forward mode scattered on any material impurities or surface imperfections. 

\begin{figure*} 
    \centering
    \includegraphics[width=12.9cm]{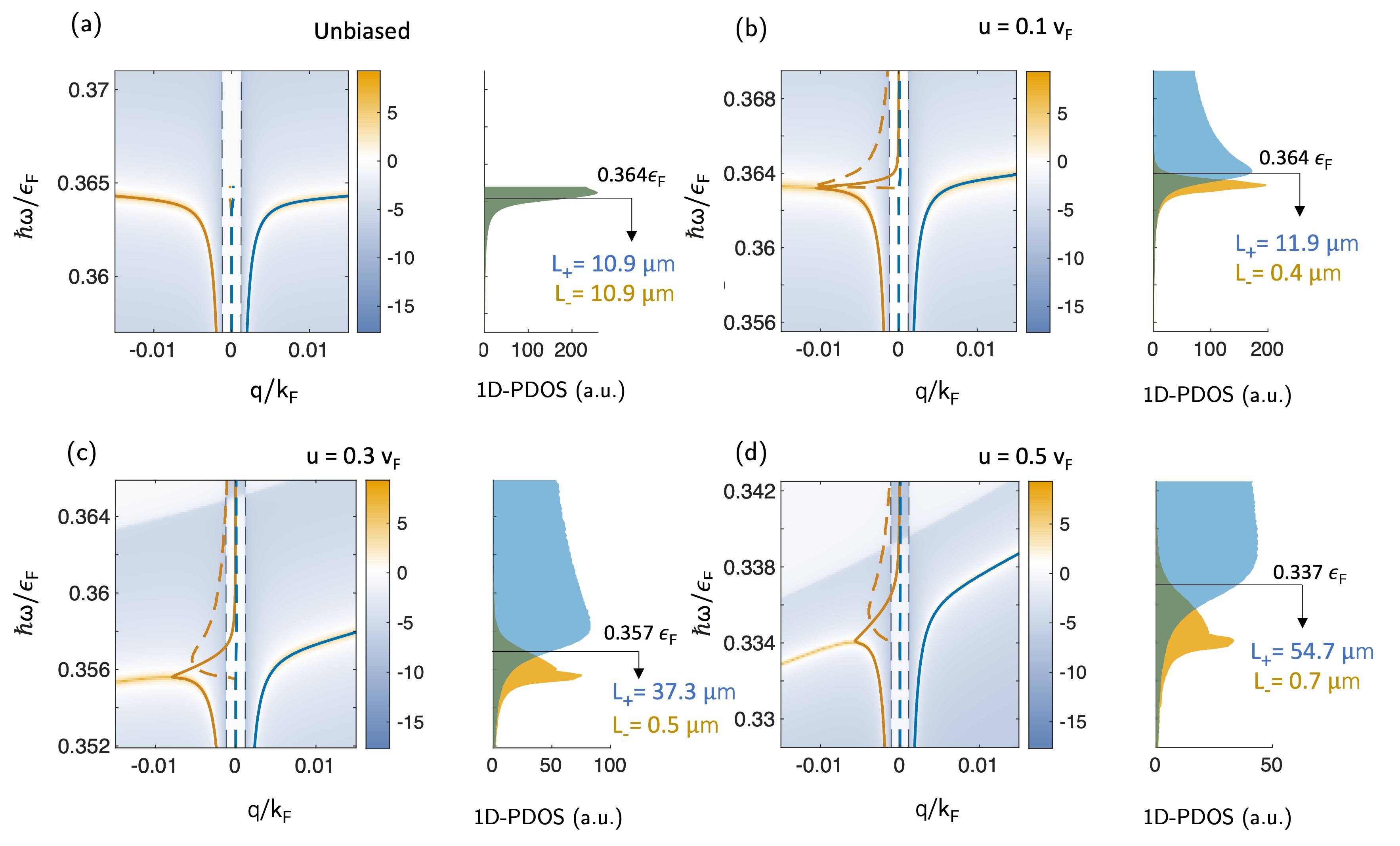}
    \caption{The SPP dispersion plots and corresponding 1D-PDOS values for the 3D DSM Cd$_3$As$_2$, which is either (a) unbiased or DC-biased with the increasing drift current velocities of (b) $u=0.1v_F$, (c) $u=0.3v_F$, and (d) $u=0.5v_F$, plotted against a heatmap of the loss function $\mbox{Im}~r_{pp}$ in log units. The propagation length is called out at a specific energy where both the relative propagation length and 1D-PDOS of the co-propagating mode (blue) dominate over the corresponding parameters of the counter-propagating mode (yellow). The same parameters for Cd$_3$As$_2$ are used as in Fig.~\ref{fig:fizeauResults} with $\tau =2.1\times 10^{-10}$s.
    }
    \label{fig:lossy-Cd3As2-DOS}
\end{figure*}

To probe the effect of progressively increasing dissipative losses on the SPP transport, we model the dispersion across the scattering period range $\tau = 2.1 \times 10^{-10} - 2.1\times10^{-12}s $\cite{Gorbar2021ElectronicSemimetals}. These data are plotted in Figs.~\ref{fig:lossy-result}(d-f) and show that Fizeau drag increases the nonlocality in the co-propagating mode, while it makes the counter-propagating mode act more local, resulting in the former being underdamped while the latter is overdamped. 
At frequencies above the -q mode inflection point in (d-f), while the counter-propagating mode is still an available channel for the optical energy dissipation via absorption, its overdamped nature translates to both low 1D-PDOS values and negligible propagation lengths. This effectively prevents SPP backpropagation, opening a window of pseudo-unidirectional surface mode transport despite the nonlocality.  This is a key result of our work, showing that Fizeau drag in 3D WSM/DSMs may unlock practical windows of pseudo-unidirectional SPP transport under real-world nonlocal and lossy conditions. At high enough scattering rates, Fig~\ref{fig:lossFunctionComparisons}(f), the co-propagating mode eventually experiences back-bending, and both modes become overdamped. 

\begin{figure*} 
    \centering
    \includegraphics[width=12.9cm]{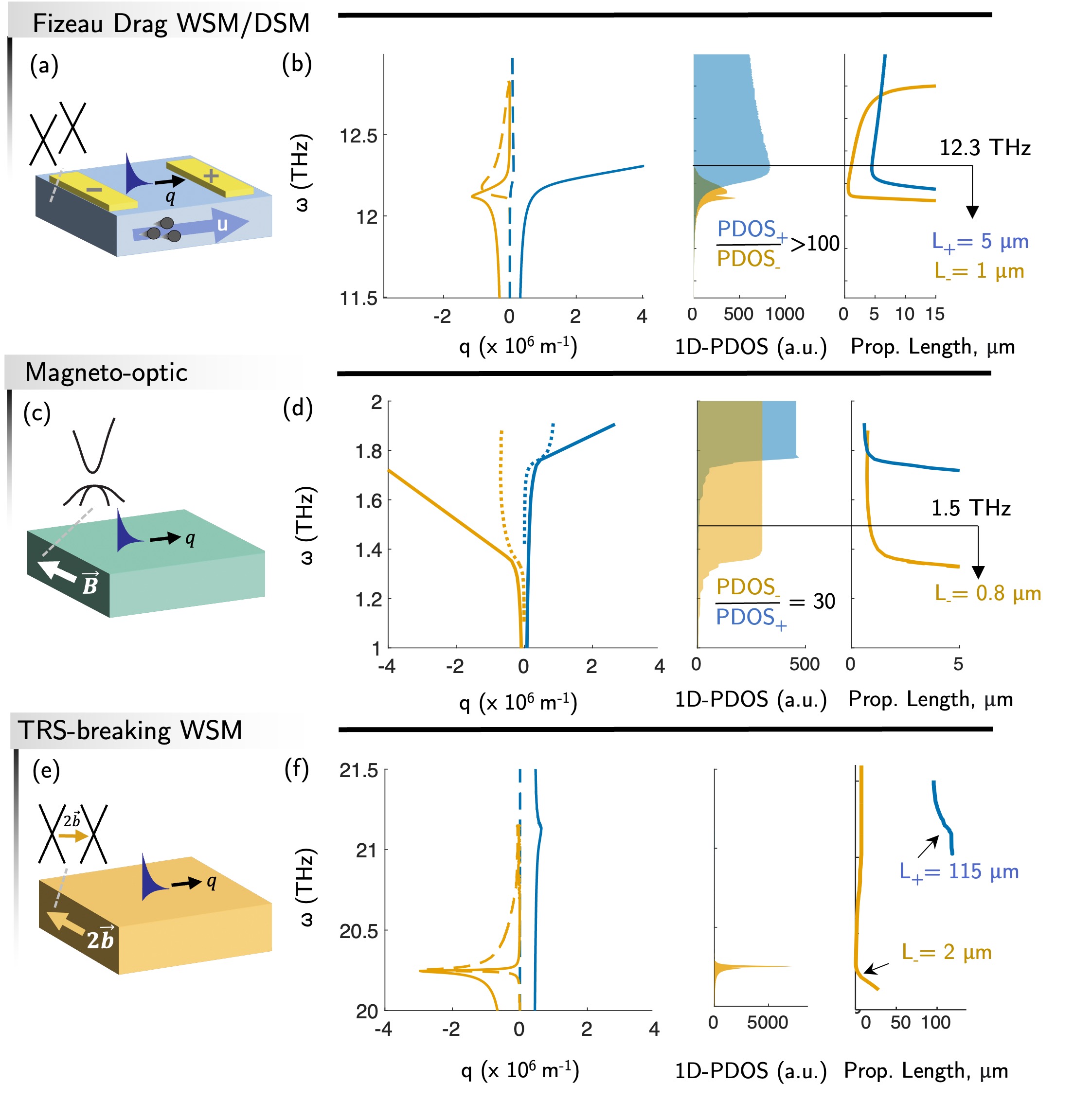}
    \caption{ \textbf{Comparison of the SPP dispersion relations and resulting 1D photonic density of states (1D-PDOS) in the nonlocal, lossy regime} at the interface of a (a-b) Fizeau drag current-biased DSM, (c-d) magneto-optic material under magnetic bias (data from Ref.~\cite{Buddhiraju2018AbsencePlasmonics}), (e-f) TRS-breaking WSM. The DC-biased WSM/DSM is modeled with the same parameters as in Fig.~\ref{fig:fizeauResults}, with  $\tau =2.1\times 10^{-11}$s and $u=0.5v_F$.
    The SPP modes on the interface of a TRS-breaking WSM and air are modeled with the commonly used parameters of $\epsilon_F = 0.15$ eV, $v_F=0.83\times10^5$ ms$^{-1}$, $\epsilon_{\infty}=13$, $\epsilon_b = 6.2$, $g=2$, a Weyl node separation of $b=2 \times 10^9$, and $\tau =2.1\times 10^{-11}$s \cite{Kotov2019GiantSemimetals, Zhao2020Axion-Field-EnabledSemimetals}. 
    The SPP modes in a magneto-optic material are shown  in Ref.~\cite{Buddhiraju2018AbsencePlasmonics} for the case of a silicon-InSb interface. The silicon is modeled with $\varepsilon=11.68$ and InSb is characterized by $\varepsilon_{\infty}=15.6$, $\omega_p=2\times 10^{12}$ Hz, $\tau = 1/0.025\omega_p=3.2\times10^{-12}$s, while an applied magnetic field is $B=0.2$T. 
    Plots (b,d,f)  have the same $q$-axes, while the $\omega$-axes have scales of 1 THz (b,d) and 1.5 THz (f) to show a relatively clear comparison in magnitude of nonreciprocity. Landau damping occurs for $q_d > \omega_P/v_F$, which is outside of the bounds of these plots for each case\cite{Boriskina2013PlasmonicApplications}. }
    \label{fig:Fiz-vs-MOs}
\end{figure*}

In Fig.~\ref{fig:lossy-Cd3As2-DOS}(a-d) we see the impact of increasing drift velocity on the pseudo-unidirectionality of SPP modes. Increased drift velocity increases the propagation length of the co-propagating mode and also modifies the ratio of available SPP states at different energies.

\subsubsection*{Comparison to nonreciprocal SPP transport in magneto-optic materials and TRS-breaking WSMs}

Similar predictions have been made for the nonreciprocity of nonlocal lossy SPP modes supported by interfaces of magneto-optic materials and time-reversal-symmetry breaking WSMs\cite{Monticone_2020, Buddhiraju2018AbsencePlasmonics}. To evaluate the advantages and pitfalls of different mechanisms of inducing nonreciprocity in 3D materials, in Fig.~\ref{fig:Fiz-vs-MOs} we compare dispersion and transport characteristics of SPPs in WSM/DSMs under Fizeau drag (a,b) with those of SPPs in magneto-optic materials under external magnetic bias (c,d) and in TRS-symmetry breaking WSMs (e,f) when loss and nonlocality are taken into account. The current-biased WSM/DSM is modeled with the same parameters as in Fig.~\ref{fig:fizeauResults}-\ref{fig:lossy-Cd3As2-DOS} with $u=0.5v_F$ and $\tau = 2.1\times 10^{-11}$s.

In comparison to the WSM/DSM Fizeau drag results, when nonlocality and loss are accounted for in the magneto-optic InSb, Fig.~\ref{fig:Fiz-vs-MOs}(c), results from Ref.~\cite{Buddhiraju2018AbsencePlasmonics} show that the forward and reverse propagating modes (1) both exhibit upward bending behavior and (2) are both underdamped, $\mbox{Im}[q]<\mbox{Re}[q]$, meaning both modes propagate. This result is replicated in Fig.~\ref{fig:Fiz-vs-MOs}(d). The 1D-PDOS plot shows that an energy window exists where the backward (yellow) mode dominates in the number of available states, $\omega\approx 1.4-1.6$ THz, opening a broad window of pseudo-unidirectional transport, demonstrated by the large 1D-PDOS$_{-}/$1D-PDOS$_{+}$ ratio. When the two modes overlap at higher $\omega$ they are in competition, exhibiting the same relative propagation length.

Next, we model the SPPs at a TRS-breaking WSM interface, Figs.~\ref{fig:Fiz-vs-MOs}(e-f), with the material parameters from Ref.~\cite{Zhao2020Axion-Field-EnabledSemimetals} and $\tau =2.1\times10^{-11}$s. We apply our semiclassical, nonlocal polarizability function, Eq.~\ref{eq:pol-normal}, to calculate the diagonal components of the permittivity. Even when accounting for nonlocality, at this magnitude of loss, the familiar back-bending behavior\cite{Pajovic2020IntrinsicSurfaces} for the $\pm q$ modes is observed. The $-q$ (yellow) mode is extremely dominant in 1D-PDOS, with the $+q$ (blue) mode 1D-PDOS being so much lower that it is not visible. However, at the peak of the $-q$ mode 1D-PDOS, the propagation length of the backward mode, $L_{-}$, is relatively low. In contrast, $L_{+}$ is much larger, but the $+q$-mode has virtually no available states in comparison to the $-q$ one. Despite this, this situation presents the most distinct pseudo-unidirectional SPP transport with the stark gap in 1D-PDOS between the  $\pm q$ modes (with the 1D-PDOS ratio $>$3000 at 20.25 THz).

Landau damping will additionally suppress the high-momentum SPP modes in Fig.~\ref{fig:Fiz-vs-MOs}, with wavevectors above the threshold of about $q_d > \omega_P / v_F$\cite{Boriskina2013PlasmonicApplications}, as discussed in Supporting Information Note~4. 
These momentum thresholds for both modes are well outside of the q-values range in Fig.~\ref{fig:Fiz-vs-MOs}. 
Surface-induced Landau damping will also impact the SPPs and is accounted for in the magneto-optic results\cite{Buddhiraju2018AbsencePlasmonics} but not for the WSM/DSM cases, Fig.~\ref{fig:Fiz-vs-MOs}(a,e). Inclusion of surface collisions would increase the scattering rates of these systems\cite{Khurgin_2015}, though the effects of nonlocality are expected to remain much more pronounced\cite{Buddhiraju2018AbsencePlasmonics}.

These results demonstrate the importance of accounting for nonlocality and loss in discussions of nonreciprocal surface plasmon-polariton transport. Additionally, they emphasize the importance of considering the 1D-PDOS$_{\pm}$ and propagation lengths, $L_{\pm}$, as key figures of merit for quantifying the directionality and efficiency of the SPP energy transport. 

\section*{Discussion}

In summary, we have presented the theoretical framework for plasmon Fizeau drag in 3D Dirac and Weyl materials, enabling it to join the toolbox of mechanisms for tuning the nonreciprocity of topological materials. Importantly for engineering applications, we have shown that the unique blend of nonlocality and loss in the DSM Cd$_3$As$_2$ enables windows of pseudo-unidirectional transport. 

Further, we have introduced new FoMs for evaluating and ranking nonreciprocal and unidirectional SPP transport: 1D-PDOS$_{\pm}$ and propagation length, $L_{\pm}$. These FoMs allow us to compare the potential for pseudo-unidirectional transport in Fizeau drag-biased 3D WSM/DSMs to that of magneto-optic materials and TRS-breaking WSMs. In terms of practicality, the magneto-optic case in Figs~\ref{fig:Fiz-vs-MOs}(c-d) requires a modest applied magnetic field of B=0.2T and a nonreciprocal absorber/emitter was recently demonstrated with the magneto-optic InAs under the external magnetic field of 1T\cite{Shayegan_Biswas_Zhao_Fan_Atwater_2023}. In contrast, the first TRS-breaking WSM was only experimentally demonstrated recently and, to date, the prediction of nonreciprocal optical behavior in these materials is in the early stages of experimental demonstration\cite{Guo2023LightSemimetals, Han_Markou_Stensberg_Sun_Felser_Wu_2022}. The experimental demonstration of Fizeau drag in 3D WSM/DSM will not require an externally applied magnetic field or corresponding equipment and will also draw on a wider selection of I-symmetry breaking DSM and WSM material candidates that may be less vulnerable to the synthesis challenges involved with working with TRS-breaking WSM materials. 

In this vein, we propose that experiments to demonstrate Fizeau drag biased 3D WSM/DSMs be pursued. 
In experiment, we expect that, as was observed in graphene\cite{Dong2021FizeauPlasmonics}, the DC bias may modify the higher order, nonlinear optical conductivity components, leading to a nonlinear Fizeau shift as a function of the DC bias field. 
Temperature should be carefully considered for experimental demonstration of this effect, as the high values of the charge carrier mobility, such as those shown in Fig.~\ref{fig:lossy-result} for Cd$_3$As$_2$, have been measured at low temperatures\cite{Liang2014UltrahighSemimetalCd3As2}. Additionally, the optical model derived here assumes low temperature; the elevated temperature will cause smearing of the Fermi level, which will reduce the Fizeau drag effect. Experimental demonstrations of Fizeau drag in graphene were conducted at cryogenic conditions and while some Joule heating was expected to raise the device temperature, the experimental results deviated minimally from the theoretical predictions. The impact of Joule heating should be considered in 3D Dirac and Weyl materials, and mitigation techniques such as heat sinks, pulsed rather than constant electrical currents, and device geometry should be examined.
Additionally, plasmon Fizeau drag in thin film DSM/WSM configurations rather than bulk slabs should reduce the effects of damping\cite{Kotov2016DielectricFilms}. 

In contrast to graphene Fizeau drag experiments, the effect in 3D WSM/DSM material platforms should open windows of pseudo-unidirectional SPP transport, allowing for additional experimental probes beyond measuring SPP wavelength. In particular, the frequency and angular selectivity of THz absorbtance or thermal emittance both scale with the 1D-PDOS values, allowing to probe nonreciprocity via far-field spectroscopic or ellipsometric measurements, similar to the experiments performed with magneto-optic materials under magnetic bias\cite{ Chochol_Postava_Čada_Pištora_2017, Shayegan_Zhao_Kim_Fan_Atwater_2022}.

We have also related the PGE to Fizeau drag as a new mechanism for ultrafast THz-driven control of quantum materials. We believe that future APRES experiments of the CPGE and LPGE in WSM/DSMs will help to map the transient electron distribution under this effect and will enable future extensions of the modeling presented here for the DC-biased Fizeau drag. The same material evaluated for the drift current case, Cd$_3$As$_2$, is an ideal candidate for CPGE-enabled Fizeau drag under strain \cite{Liang2014UltrahighSemimetalCd3As2, neupane_observation_2014, Shoron_Schumann_Goyal_Kealhofer_Stemmer_2019}.
Additionally, the Fizeau drag effect may be extended to surface PGE currents as well as surface currents in TIs.  

Finally, it was recently experimentally shown that photocurrents in type-I WSM TaAs (identified to have large $v_F$ in Fig.~\ref{fig:fizeau}(e)) can break TRS in the material\cite{Sirica2021Photocurrent-drivenTaAs}. The degree of TRS-breaking was directly related to the direction of photocurrent\cite{Sirica2022UsingTaAs} and it was deduced to be entirely electronic in origin.  We posit that the electronic origin of this TRS-breaking mechanism may be understood and quantified through the Fizeau drag mechanism presented here.

\begin{acknowledgments}
We wish to acknowledge the support of Draper Labs for funding M.G.B. through the Draper Scholars Program. This work is supported in part by the Near Field Radiative Heat Transfer ARO MURI (Grant No. W911NF-19-1-0279) via U. Michigan. We thank Yoichiro Tsurimaki, Shanhui Fan, Thanh Nguyen, Simo Pajovic, and  Mark Witinski for helpful discussions. 
\end{acknowledgments}

\medskip

\noindent \textbf{Supporting Information Available:} This contains a nomenclature table, (I) a reproduction of the RPA polarizability model of doped Dirac/Weyl semimetals, (II) a detailed semiclassical derivation of the polarizability and permittivity of doped Dirac/Weyl semimetals in the nonlocal regime, (III) a detailed semiclassical derivation of the polarizability and permittivity of drift current biased doped Dirac/Weyl semimetals in the nonlocal regime, and (IV) a discussion of the complex-q solving method used to find the surface plasmon modes as well as discussions of collision-induced dissipative losses and Landau damping in plasmon Fizeau drag biased Weyl/Dirac semimetals.


\bibliography{apssamp}

\clearpage
\onecolumngrid
\begin{center}
\textbf{\large Supplementary Information for ``Plasmon Fizeau drag in 3D Dirac and Weyl semimetals"\\}
\bigbreak
Blevins and Boriskina
\end{center}
\setcounter{equation}{0}
\setcounter{figure}{0}
\setcounter{table}{0}
\setcounter{page}{1}
\setcounter{section}{0}
\makeatletter
\renewcommand{\theequation}{S\arabic{equation}}
\renewcommand{\thefigure}{S\arabic{figure}}
\renewcommand{\thetable}{S\arabic{table}}


\begin{table}[h]
\caption{\label{tab:table1}%
Nomenclature
}
\begin{ruledtabular}
\begin{tabular}{@{}ll@{}}
\toprule
\textbf{Acronyms}         &                                                                                   \\ \midrule
DSM                       & Dirac Semimetal                                                                   \\
LPGE                      & Linear Photogalvanic Effect                                                       \\
QLT                       & Quasi-Lorentz Transform                                                           \\
RPA                       & Random Phase Approximation                                                        \\
SC                        & Semiclassical                                                                     \\
SPP                       & Surface Plasmon Polariton                                                         \\
THz                       & Terahertz                                                                         \\
WSM                       & Weyl Semimetal                                                                    \\
\textbf{Electromagnetics} &                                                                                   \\ \midrule
$\omega$                  & EM frequency                                                                      \\
$\omega_0$                & EM frequency in moving frame                                                      \\
$\omega_p$                & EM plasma frequency                                                               \\
$\sigma$                  & Optical conductivity                                                              \\
$\sigma^u$                & Optical conductivity for current-biased state                                     \\
$\varepsilon$             & Permittivity                                                                      \\
$\varepsilon_\infty$   & Background permittivity                                                           \\
$q$                       & $ = 2 \pi/\lambda $, Wavevector                                                   \\
$q_0$                     & Wavevector in moving frame                                                        \\
$q_d$                     & Landau damping wavevector                                                         \\
\textbf{Matter/Electrons} &                                                                                   \\ \midrule
$\epsilon$                & Energy                                                                            \\
$\epsilon_F$              & Fermi Energy                                                                      \\
$\gamma$                  & $=(1-u^2/v_F^2)^{-1/2}$ Quasi-Lorentz factor                                      \\
$\mu$                     & Chemical potential                                                                \\
$\mu^u$                   & Transformed chemical potential in current-carrying state                          \\
$\mu_0$                   & Chemical potential in moving frame                                                \\
$\phi$                    & Potential                                                                         \\
$\Pi^u_{\rho,\rho}$       & $=\Pi^u$, Polarizability, density-density response function,   under current bias \\
$\Pi_{\rho,\rho}$         & $=\Pi$, Polarizability, density-density response function                         \\
$\rho_{ind}$              & Induced charge density                                                            \\
$k_F$                     & Fermi momentum                                                                    \\
$n$                       & Quasiparticle density                                                             \\
$N(\epsilon)$             & Density of states                                                                 \\
$n_0$                     & Quasiparticle density in moving frame                                             \\
$p$                       & Quasiparticle momentum                                                            \\
$p_0$                     & Quasiparticle momentum in moving frame                                            \\
$p_F$                     & Fermi momentum                                                                    \\
$u$                       & Carrier drift velocity                                                            \\
$v_F$                     & Fermi velocity                                                                    \\
$g$                       & Degeneracy                                                                        \\
$g_W$                     & $= g/2$ Number of pairs of Weyl nodes                                             \\ \bottomrule
\end{tabular}
\end{ruledtabular}
\end{table}

\section*{Supplemental Note 1:  RPA longitudinal polarizability in DSM and WSM materials} \label{sec:A-RPA}

 The regular, unbiased longitudinal dynamical polarizability, $\Pi(\bf{q},\omega)$ (also called the density-density response $\Pi_{\rho,\rho}(\bf{q},\omega)$, although we drop the $\rho,\rho$ subscript), and the resulting dielectric function, $\varepsilon(\mathbf{q},\omega)$, of doped WSM/DSMs have been derived in the literature via the first-principles linear response theory in the random phase approximation (RPA)~\cite{Thakur2017DynamicalSystems,Thakur2018DynamicSemimetals, Lv_Zhang_2013, Kotov2016DielectricFilms} and is reproduced below for comparison with our semiclassical, nonlocal result in Supporting Information Note 2.
\begin{equation} \label{eq:pol-RPA}
\begin{aligned}
    \mbox{Re}\Pi^{RPA}({{q}}, \omega) &= -\frac{gq^2}{24 \pi^2 \hbar v_F} \ln{\frac{4 v_F^2 q_{max}^2}{|v_F^2 q^2 - \omega^2|}} - \frac{g q^2}{8 \pi^2 \hbar v_F}[C(q,\omega)+D(q,\omega)], \\
    \mbox{Im}\Pi^{RPA}({{q}}, \omega) 
    & =  -\frac{gq^2}{24 \pi \hbar v_F} \Theta(\omega-v_Fq) - \frac{g}{8 \pi \hbar v_F} 
    \begin{cases}
      \zeta(q,\omega) - \zeta(q,-\omega),  & \text{1A}\\
      \zeta(q,\omega),  & \text{2A}\\
      -\frac{q^2}{3}, & \text{1B} \\
      - \zeta(-q,-\omega),  & \text{2B}\\
      0, & \text{3A,3B}
    \end{cases} \\
\end{aligned}
\end{equation}

\noindent where $\mathbf{q}=q \mathbf{\hat{z}}$, $q_{max}=\epsilon_{max}/\hbar v_F$ is the cutoff wavevector corresponding to UV energy cutoff $\epsilon_{max}$, and $g$ is the degeneracy (for WSMs, $g=2g_W$, where $g_W$ is the number of pairs of Weyl nodes). The domains in Eq.~\ref{eq:pol-RPA} are defined as 

\begin{equation} \label{eq:pol-RPA-domains}
\begin{aligned}
    \mbox{1A}:~~ &0<\omega<v_Fq &\mbox{and}~~ &2\mu - \hbar v_F q - \hbar \omega >0, \\
    \mbox{2A}:~~ &0<\omega<v_Fq &\mbox{and}~~ &\pm 2\mu \mp \hbar v_F q + \hbar \omega >0, \\
    \mbox{3A}:~~ &0<\omega<v_Fq &\mbox{and}~~ & 2\mu - \hbar v_F q + \hbar \omega < 0, \\
    \mbox{1B}:~~ &0<v_F q <\omega &\mbox{and}~~ & 2\mu - \hbar v_F q - \hbar \omega > 0, \\
    \mbox{2B}:~~ &0<v_F q <\omega &\mbox{and}~~ & \mp 2\mu \pm \hbar \omega + \hbar v_F q > 0, \\
    \mbox{3B}:~~ &0<v_F q <\omega &\mbox{and}~~ & 2\mu + \hbar v_F q - \hbar \omega < 0, \\
\end{aligned}
\end{equation}

\noindent and the defined functions are

\begin{equation} \label{eq:pol-RPA-funcs}
\begin{aligned}
    \zeta(q,\omega) &= \frac{1}{12 \hbar^3 v_f^3 q} [(2\mu + \hbar \omega)^3 - 3\hbar^2 v_F^2 q^2 (2\mu +\hbar\omega) + 2\hbar^3 v_F^3 q^3],\\
    C(q,\omega) &= \frac{8 \mu^2}{3 \hbar^2 v_F^2 q^2} - \frac{\zeta(q,\omega) H(q,\omega)}{q^2} - \frac{\zeta(-q,\omega) H(-q,\omega)}{q^2}, \\
    D(q,\omega) &= C(q,-\omega) - \frac{8 \mu^2}{3 \hbar^2 v_F^2 q^2},~~
    H(q,\omega) = \ln{ \frac{|2\mu +\hbar \omega - \hbar v_Fq|}{|\hbar v_F q -\hbar \omega|} }, \\
\end{aligned}
\end{equation}

\noindent where $\mu$ is the chemical potential. Note that chemical potential is temperature dependent, $\mu(\epsilon_F, T)$, and at zero temperature is equal to the Fermi energy $\epsilon_F$.

\section*{Supplemental Note 2: Semiclassical longitudinal polarizability and optical conductivity in DSM and WSM materials}\label{secA1}

\subsection*{Nonlocal Case}
To enable the derivation of a Fizeau drag-modified optical conductivity for 3D DSM/WSMs, we first derive the nonlocal Drude conductivity of the 3D DSM/WSMs via a semiclassical Boltzmann equation formalism ~\cite{Ziman1972PrinciplesSolids, Taylor2002SemiclassicalMetals}. The quasiparticle distribution function, $f(\mathbf{ p}) = f^0(\mathbf{ p}) + \delta f(\mathbf{ p})$, is modeled as a combination of the equilibrium distribution, $f^0(\mathbf{ p})$, and a perturbation to the distribution, $ \delta f(\mathbf{ p})$, via an externally applied electric field, $\mathbf{E}= \mathbf{E_0}e^{-i\omega t + i \mathbf{q}\cdot \mathbf{r}}$.  We will use the collisionless regime, $(\partial f/\partial t)_{coll} \rightarrow 0$, and account for losses afterward. The Boltzmann transport equation is

\begin{equation}
    \begin{aligned}
        \frac{\partial f}{\partial t} + \dot{\mathbf{ p}} \cdot \nabla_p f + \mathbf{ v_p} \cdot \nabla_r f  &= 0 .\\
    \end{aligned}
\end{equation}

\noindent Making use of $\dot{\mathbf{ p}} = e \nabla_r \phi_{tot}(\mathbf{ r},t)$ and $\nabla_p f^{0}(\mathbf{ p}) = \nabla_{\mathbf{ p}}\epsilon_p \frac{\partial f^0(\epsilon_p)}{\partial \epsilon_p} = \mathbf{ v_p}\frac{\partial f^0(\epsilon_p)}{\partial \epsilon_p}$ and after several simplifying steps, the perturbed distribution is~\cite{Bruus2004FermiTheory}

\begin{equation}
    \begin{aligned}
        \delta f &= \frac{\mathbf{ q} \cdot \mathbf{ v_p}} {\omega -\mathbf{ q} \cdot \mathbf{ v_p} } \left(-\frac{\partial f^0(\epsilon_p)}{\partial \epsilon_p} \right)[-e \phi_{tot}(\mathbf{ q},\omega)], \\
    \end{aligned}
\end{equation}
and the induced charge density is subsequently

\begin{equation}
\begin{aligned}
    \rho_{ind}(\mathbf{ q}, \omega) &= \frac{g}{\mathcal{V}} \sum_\mathbf{ k} \frac{\mathbf{ q} \cdot \mathbf{ v_p}} {\omega -\mathbf{ q} \cdot \mathbf{ v_p} } \left(-\frac{\partial f^0(\epsilon_p)}{\partial \epsilon_p}\right)[-e \phi_{tot}(\mathbf{ q},\omega)] , \\
\end{aligned}
\end{equation}
where $\mathcal{V}$ is the normalization volume and $g$ is the degeneracy. From the induced charge density, we can extract the dynamical polarizability function, also called the density-density response function, $\Pi_{\rho ,\rho}(\mathbf{q},\omega) = \rho_{ind}/(-e\phi_{tot})$  as~\cite{Mahan_2000}

\begin{equation}
\begin{aligned}
    \Pi_{\rho,\rho}(\mathbf{ q}, \omega) & = - \frac{g}{\mathcal{V}} \sum_\mathbf{ p} \frac{\mathbf{ q} \cdot \mathbf{ v_p}}{\omega- \mathbf{ v_p \cdot q}} \frac{\partial f^0(\epsilon_p)}{\partial \epsilon_p} .\\ 
\end{aligned}
\end{equation}

\noindent Note that we are dropping the $\rho, \rho$ subscript for the rest of the text, using $\Pi_{\rho ,\rho}(\mathbf{q},\omega)=\Pi(\mathbf{q},\omega)$. The integral can then be transformed from 3-dimensional momentum-space to the one over the solid angle, $\Omega$, and energy~\cite{Coleman2015Finite-temperaturePhysics,Borgnia2015Quasi-RelativisticGraphene}: 

\begin{equation} \label{eq:solid-angle-transform}
\begin{aligned}
\frac{1}{\mathcal{V}} \sum_\mathbf{ p} &\rightarrow \int \frac{d\Omega}{4\pi} d\epsilon N(\epsilon) \rightarrow \int_0^{2\pi} d\phi \int_0^{\pi} \frac{d\theta sin\theta}{4\pi} \int d\epsilon N(\epsilon) ,
 \\
\end{aligned}
\end{equation}

\noindent where $N(\epsilon) = \frac{\epsilon^2}{2\pi^2\hbar^3v_F^3}$ is the electron density of states for a WSM/DSM.
Thus,
\begin{equation}
\begin{aligned}
    \Pi(\mathbf{ q}, \omega) & = - g  \int_0^{2\pi} d\phi \int_0^{\pi} \frac{d\theta sin\theta}{4\pi}  \frac{\mathbf{ q} \cdot \mathbf{ v_p}}{\omega- \mathbf{ v_p \cdot q}} \int d\epsilon N(\epsilon) \frac{\partial f^0(\epsilon_p)}{\partial \epsilon_p}. \\ 
\end{aligned}
\end{equation}

\noindent Note that for low temperature, $kT \ll \epsilon_F$ $, \partial_{\epsilon}f^{0} \approx -\delta \left( \epsilon (\mathbf{ p})-\mu \right) $ ~\cite{Taylor2002SemiclassicalMetals}, leading to:
\begin{equation}
\begin{aligned}
    \Pi(\mathbf{ q}, \omega) & = g  N(\mu)\int_0^{2\pi} d\phi \int_0^{\pi} \frac{d\theta sin\theta}{4\pi}  \frac{\mathbf{ q} \cdot \mathbf{ v_p}}{\omega- \mathbf{ v_p \cdot q}}.  \\ 
\end{aligned}
\end{equation}

To solve for the longitudinal polarizability, $\Pi_{\rho_z,\rho_z}({q}, \omega)$, we will consider the case where the electric field, $\mathbf{E}$, is parallel to the direction of propagation, $\mathbf{E} ||  \mathbf{q} || \mathbf{\hat{z}}$. Moving forward we will again drop the $\rho,\rho$ subscript and use $\Pi({q}, \omega)$. Note $\nabla_p\epsilon = \mathbf{v_p} = v_F (sin{\theta}\cos{\phi}$, $\sin\theta\sin\phi$, $\cos{\theta})$, thus $v_z = v_F\cos{\theta} $. We assume an isotropic electron band structure, specifically an isotropic Fermi surface such that $v_F\hat{x} =v_F\hat{y} = v_F\hat{z} $.  This assumption has been used in the RPA for WSM/DSM permittivity and the implications of anisotropic bandstructure are discussed in Ref.~\cite{Thakur2018DynamicSemimetals}. This assumption allows us to consider $\mathbf{q} = q\hat{z}$ for simplicity of integration: 

\begin{equation}
\begin{aligned}
    \Pi({q}, \omega) &= g N(\mu) \frac{1}{4\pi}  \int_0^{2\pi} d\phi \int_0^{\pi} d\theta \sin\theta\frac{{q}{v_F \cos\theta}}{\omega- {qv_F \cos\theta}}.   \\ 
\end{aligned}
\end{equation}

\noindent Let $a=\frac{qv_F}{\omega}\cos\theta$ , $\frac{da}{d\theta} = -\frac{qv_F}{\omega}\sin\theta$, $d\theta \sin\theta = -da \frac{\omega}{q v_F}$. Then,
\begin{equation} \label{eq:local-v-nonlocal-spot}
\begin{aligned}
    \Pi({ q}, \omega) 
    &= g N(\mu) \frac{1}{4\pi}   \int_0^{2\pi} d\phi \int_{qv_F/\omega}^{-qv_F/\omega} \left(\frac{-\omega}{qv_F} \right)da  \frac{a}{1-a},   \\
    &= g N(\mu) \frac{1}{4\pi}   \left(\frac{\omega}{qv_F} \right) {2\pi}   \left[ a + \ln{(1-a)} \right]_{qv_F/\omega}^{-qv_F/\omega}. \\
\end{aligned}
\end{equation}

\noindent Assuming a complex-valued logarithm, this solves to

\begin{equation} \label{eq:NL-pol}
\begin{aligned}
    \Pi({q}, \omega) 
    &= - g N(\mu) \left(\frac{\omega}{qv_F}\frac{1}{2}\ln{\frac{\omega-qv_F}{\omega+qv_F}} + 1 \right). \\
\end{aligned}
\end{equation}

This is the semiclassical, nonlocal dynamical polarizability function as presented in the main text, $\Pi^{SC,NL}({q}, \omega)$ (Eq.~\ref{eq:pol-normal}). For thoroughness, we define the real and imaginary parts

\begin{equation} \label{eq:NL-pol-Im-Re}
\begin{aligned}
    \mbox{Re}~\Pi({q}, \omega) 
    &= \mbox{Re} \left[- g \frac{\mu^2}{2\pi^2\hbar^3v_F^3} \left(\frac{\omega}{qv_F}\frac{1}{2}\ln{\frac{\omega-qv_F}{\omega+qv_F}} + 1 \right) \right], \\
    \mbox{Im}~\Pi({q}, \omega) 
    &= \mbox{Im} \left[- g \frac{\mu^2}{2\pi^2\hbar^3v_F^3} \left(\frac{\omega}{qv_F}\frac{1}{2}\ln{\frac{\omega-qv_F}{\omega+qv_F}} + 1 \right) \right], \\
    & = 0,~~\omega>v_F q.
\end{aligned}
\end{equation}

In the semiclassical regime, it is assumed that the energy scales of $\hbar \omega$ and $v_F q$ are much less than $\mu$, so the model here is compared to the RPA result~\cite{Thakur2018DynamicSemimetals} over the $2\mu-\hbar v_Fq-\hbar \omega>0$ domain in the main text Fig.~\ref{fig:pol-func}. Qualitatively, the semiclassical model shows good agreement with the RPA model.

\begin{figure}
    \centering
    \includegraphics[width=1\linewidth]{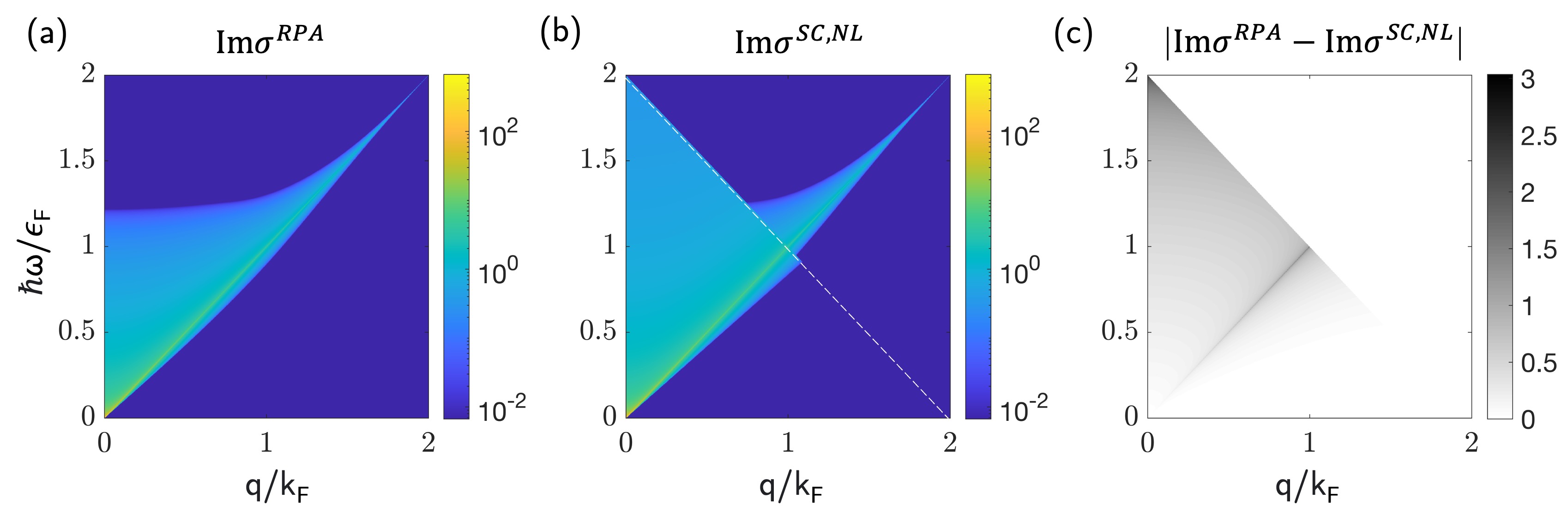}
    \caption{\textbf{The imaginary part of the optical conductivity, $\mbox{Im}\sigma$}, from the RPA model (a) and from the semiclassical nonlocal (SC,NL) derivation in this work (b). In (b), the semiclassical $\sigma^{SC,NL}$ is plotted on the $2\mu-\hbar v_Fq-\hbar \omega>0$ domain, corresponding to the area below the white dashed line, and the RPA solution is  kept for comparison for $2\mu-\hbar v_Fq-\hbar \omega<0$ . In (c), the difference between the RPA and semiclassical $\mbox{Im}\sigma$ is plotted. All plots are in units of $e^2 g k_F / 4 \pi^2 \hbar $.  }
    \label{fig:cond}
\end{figure}

The longitudinal polarizability can then be converted to longitudinal optical conductivity, in this case $\sigma_{zz}$, via ~\cite{Bruus_Flensberg_2004}
\begin{equation} \label{eq:Normal-cond-Im}
\begin{aligned}
    {\mbox{Im}\sigma_{zz}} (q,\omega) & = \frac{ie^2\omega}{q^2} \mbox{Re}\Pi (q, \omega), \\ 
    & = \frac{ie^2\omega}{q^2} \mbox{Re}\left[- g N(\mu) \left(\frac{\omega}{qv_F}\frac{1}{2}\ln{\frac{\omega-qv_F}{\omega+qv_F}} + 1 \right) \right], \\ 
\end{aligned}
\end{equation}
\begin{equation} \label{eq:Normal-cond-Re}
\begin{aligned}
    {\mbox{Re}\sigma_{zz}} (q,\omega) & = -\frac{e^2\omega}{q^2} \mbox{Im}\Pi (q, \omega), \\ 
    & = \frac{e^2\omega}{q^2} \mbox{Im}\left[g N(\mu) \left(\frac{\omega}{qv_F}\frac{1}{2}\ln{\frac{\omega-qv_F}{\omega+qv_F}} + 1 \right) \right]. \\ 
\end{aligned}
\end{equation}
As shown in Supporting Information Fig.~\ref{fig:cond}, the resulting optical conductivity in the semiclassical model aligns with the RPA model best in the region $2\mu-\hbar v_Fq-\hbar \omega>0$ and $ \omega < v_F q$ . As expected, the semiclassical model deviates from the RPA result at higher energy, $ \omega > v_F q$.

\begin{figure} 
    \centering
    \includegraphics[width=0.35\linewidth]{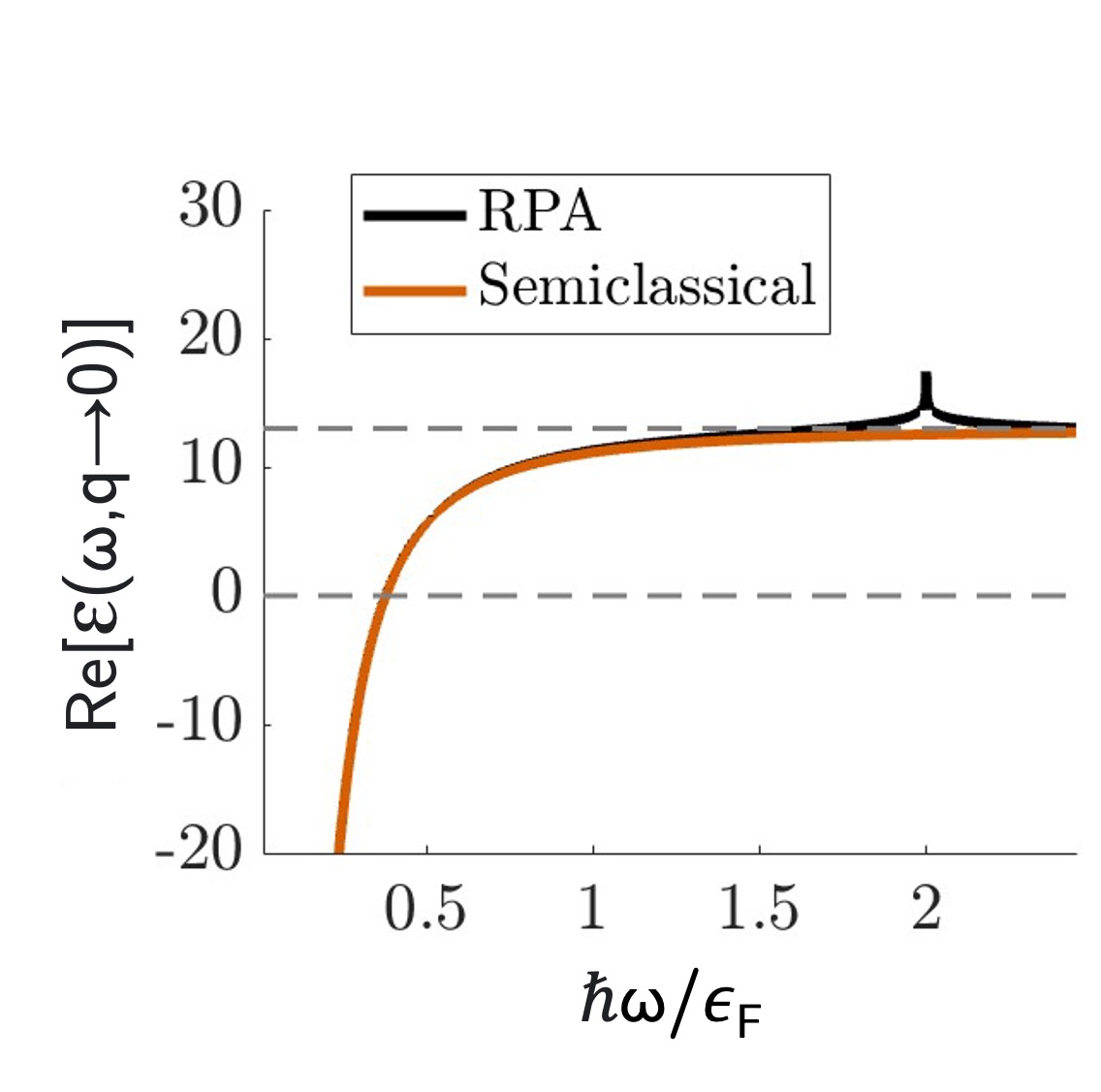}
    \caption{The real part of the WSM/DSM permittivity in the local approximation, $\varepsilon(\omega)$, for the case of $\epsilon_F = 0.15$ eV,  $v_F = 1 \times 10^6$ m/s, $g=4$, $\varepsilon_\infty=13$, and $\varepsilon_b=12$ characteristic to Cd$_3$As$_2$~\cite{Kotov2016DielectricFilms}. In the RPA $\epsilon_{max} = 3 \epsilon_F$.}
    \label{fig:permittivity}
\end{figure}

Finally, recall that the semiclassical $\Pi(q,\omega)$ derived here accounts for only intra-band transitions. The material permittivity, $\varepsilon(q,\omega)$, accounting for just intra-band transitions is found via~\cite{Kotov2016DielectricFilms,Dey2022DynamicalSystem,Thakur2017DynamicalSystems,Thakur2018DynamicSemimetals,Zhao2020Axion-Field-EnabledSemimetals}

\begin{equation}
\begin{aligned}
        \varepsilon(q, \omega) 
        &= \varepsilon_\infty (1 - V_q \Pi^{intra}(q,\omega)), \\
        &= \varepsilon_\infty + i \frac{\sigma^{intra}}{\varepsilon_0 \omega},\\
\end{aligned}
\end{equation}

\noindent where $\varepsilon_{\infty}$ is the background permittivity and $V_q = e^2/(\varepsilon_\infty \varepsilon_0 q^2)$ is the Fourier transform of the Coulomb potential energy in SI units~\cite{Dey2022DynamicalSystem}. The resulting local permittivity values ($q \rightarrow 0$) of the semiclassical and RPA approaches show good agreement (Fig.~\ref{fig:permittivity}(a)): in the usual Drude fashion, the semiclassical Drude result lacks the peak at $\hbar\omega/\epsilon_F=2$, which corresponds to inter-band absorption~\cite{Kotov2016DielectricFilms}.  

Note that for combined inter- and intra-band contributions, the relationship between $\Pi(q, \omega)$ and $\varepsilon (q,\omega)$ is
\begin{equation}
\begin{aligned}
        \varepsilon(q,\omega) 
        &= \varepsilon_b (1 - V_q (\Pi^{intra}(q,\omega)+\Pi^{inter}(q,\omega))), \\
        &= \varepsilon_b + i \frac{(\sigma^{intra}(q,\omega)+\sigma^{inter}(q,\omega))}{\varepsilon_0 \omega},\\ \\
\end{aligned}
\end{equation}
where $\varepsilon_b$ is a function of $\varepsilon_\infty$~\cite{Kotov2016DielectricFilms}.

Finally, the effects of loss via scattering are accounted for in the polarizability function by introducing a phenomenological relaxation time parameter $\tau$, which accounts for losses due to electron scattering on impurities and lattice defects as well as electron-phonon scattering~\cite{JablanPlasmonicsFrequencies,Lin2017All-angleHeterostructures}:
\begin{equation} 
\begin{aligned}
    \Pi_{\tau}({q}, \omega) 
    & = \frac{(1+i/\omega\tau)\Pi(q,\omega+i/\tau)}{1+(i/\omega\tau)\Pi(q,\omega+i/\tau)/\Pi(q,0)}.  \\
\end{aligned}
\end{equation}

\noindent Thus, 
\begin{equation} 
\begin{aligned}
    \Pi_{\tau}({q}, \omega) 
    & = \frac{(1+i/\omega\tau)g N(\mu) \left(-\frac{\omega+i/\tau}{qv_F}\frac{1}{2}\ln{\frac{\omega+i/\tau+qv_F}{\omega+i/\tau-qv_F}} +1 \right)}{1+(i/\omega\tau) \left( -\frac{\omega+i/\tau}{qv_F}\frac{1}{2}\ln{\frac{\omega+i/\tau+qv_F}{\omega+i/\tau-qv_F}} +1 \right)} . \\
\end{aligned}
\end{equation}

\subsection*{Local model}

For completeness, we show that this same approach leads to the well-known local Drude approximation for WSM/DSMs when $\omega \gg qv$ is assumed. The nonlocal semiclassical model can be reduced to the local one by using the Taylor expansion of the polarizability function  in powers of $q$ around $q\rightarrow 0$~\cite{Bruus_Flensberg_2004}. Starting from Eq.~\ref{eq:local-v-nonlocal-spot}, we obtain:

\begin{equation} \label{eq:polar-xx-c}
\begin{aligned}
    \frac{a}{1-a} \approx a + a^2 + O[a]^3 . \\ 
\end{aligned}
\end{equation}
Inserting this term into the expression for the polarizability, we arrive at the local semiclassical model:
\begin{equation} \label{eq:polar-xx-e}
\begin{aligned}
    \Pi({q}, \omega) & \approx g N(\mu) \frac{1}{4\pi}   \left(\frac{\omega}{qv_F} \right) {2\pi}   \left[ \frac{1}{2}a^2 + \frac{1}{3}a^3 \right]_{-qv_F/\omega}^{qv_F/\omega} ,  \\ 
    & = \frac{1}{3} g N(\mu)\frac{q^2v_F^2}{\omega^2}   =  \frac{1}{6} \frac{{q}^2 g}{\omega^2} \frac{\mu^2}{\pi^2\hbar^3v_F}. \\
\end{aligned}
\end{equation}

\noindent The longitudinal polarizability is converted to longitudinal optical conductivity: 
\begin{equation}
\begin{aligned}
    \mbox{Im}~{\sigma_{zz}} (q,\omega) 
    & = i\frac{1}{6} \frac{ge^2}{\omega} \frac{v_F k_F^2}{\pi^2 \hbar}.
\end{aligned}
\end{equation}

\noindent This is equivalent to the known local optical conductivity for WSMs and DSMs $ \sigma_{zz} = \frac{i \varepsilon_\infty}{4 \pi \omega } \omega_p^2 $, where $\omega_p = \frac{2 k_F^2 v_F e^2 g}{3\hbar \pi\varepsilon_\infty}$~\cite{Kotov2016DielectricFilms}.
Accounting for loss, these expressions take the following form:
\begin{equation} \label{eq:polar-xx-loss}
\begin{aligned}
    \Pi_{\tau}({q}, \omega) 
    & = \frac{1}{6} \frac{{q}^2 g}{\omega(\omega+i/\tau)} \frac{\mu^2}{\pi^2\hbar^3v_F},  \\
\end{aligned}
\end{equation}

\begin{equation}\label{eq:DSM-boltz-drude}
\begin{aligned} 
        \sigma_{zz,\tau}
        &= \frac{i \varepsilon_\infty}{4 \pi (\omega+i/\tau) } \omega_p^2. \\
\end{aligned}
\end{equation}


\section*{Supplemental Note 3: Longitudinal polarizability and optical conductivity in DSMs and WSMs with current bias }\label{secA2}

\subsection*{Nonlocal case}
Here, we derive the dynamical longitudinal polarizability, $\Pi_{\rho,\rho}^u(q,\omega)$, and bulk optical conductivity, $\sigma^u(q,\omega)$, of either a Dirac semimetal or an I-symmetry breaking Weyl semimetal in a drift current-biased state. As in Supporting Information Note 2, the quasiparticle distribution function is modeled as a combination of the equilibrium distribution, $f^u(\mathbf{ p})$, and a perturbation to the distribution, $ \delta f(\mathbf{ p})$, $f(\mathbf{ p}) = f^u(\mathbf{ p}) + \delta f(\mathbf{ p})$,  assuming an externally applied electric field, $\mathbf{E}= \mathbf{E_0}e^{-i\omega t + i \mathbf{q}\cdot \mathbf{r}}$. In the case of an applied DC current, the equilibrium distribution is a skewed Fermi distribution of the form~\cite{Borgnia2015Quasi-RelativisticGraphene, Dong2021FizeauPlasmonics}:

\begin{equation}
    f^u(\mathbf{ p}) = \frac{1}{e^{(E(\mathbf{p})- \mathbf{u} \cdot \mathbf{p}- \mu^u )/k_BT} +1 },
\end{equation}

\noindent where $E(\mathbf{p})=v|\mathbf{ p}|$. We will express $\nabla_p f^{u}(\mathbf{ p})$  as an integral over the independent energy variable, $\epsilon$ \cite{Borgnia2015Quasi-RelativisticGraphene}
\begin{equation}
    \begin{aligned}
         \nabla_p f^{u}(\mathbf{ p}) = \int d \epsilon \delta(\epsilon - E(\mathbf{p})) (\mathbf{ v_p-u})\frac{\partial f^u(\epsilon, \mathbf{ p})}{\partial \epsilon},
    \end{aligned}
\end{equation}
\noindent where
\begin{equation}
    f^u(\mathbf{ p}) = \frac{1}{e^{(\epsilon - \mathbf{u} \cdot \mathbf{p}- \mu^u )/k_BT} +1 },
\end{equation}
\noindent and note that $ v_\mathbf{ p} = \frac{v^2\mathbf{ p}}{\epsilon}$. From the above equations, we can express the polarizability as

\begin{equation} \label{eq:pol1}
\begin{aligned}
    \Pi^u(\mathbf{ q}, \omega) &= - \frac{g}{\mathcal{V}} \sum_\mathbf{ p} \frac{\mathbf{ q} \cdot \nabla_p f^u(\mathbf{ p})}{\omega- \mathbf{ v_p \cdot  q}}, \\
    & = - \frac{g}{\mathcal{V}} \sum_\mathbf{ p} \int d\epsilon \delta(\epsilon - E(\mathbf{p}))\frac{ v^2\mathbf{ q \cdot  p}- \epsilon\mathbf{ q \cdot u} } {\epsilon\omega - v^2\mathbf{ p}\cdot \mathbf{q} } \frac{\partial f^u(\epsilon, \mathbf{ p})}{\partial \epsilon}. \\
\end{aligned}
\end{equation}

\noindent To solve Eq.~\ref{eq:pol1}, we will substitute, for convenience, the ansatz below for a quasi-Lorentz transformation (QLT) for the electromagnetic and electric quantities. This substitution is justified as we consider the case of the current traveling in the z-direction. We will see that this ansatz for the transformation greatly simplifies the derivation and also reveals the QLT nature of current bias in DSM/WSMs. 

\begin{equation} \label{eq:QLTs-Material}
    \epsilon_0(p)=\gamma\left(\epsilon(p)-up^z\ \right),~
    p_{0}^z=\gamma\left(p^z-\frac{u}{v_F^2}\epsilon(p)\right),~
    p_{0}^x=p^x,~p_{0}^y=p^y,
\end{equation}

\begin{equation} \label{eq:QLTs-EM}
\begin{aligned} 
        \omega_0=\gamma(\omega -u q_z ),~
        q_{0}^z&=\gamma(q^z-\frac{u}{v_F^2}\omega),~
        q_{0}^x=q^x,~
        q_{0}^y=q^y, ~
\end{aligned}
\end{equation}

\begin{equation} \label{eq:QLT-factor}
    \gamma = (1 - u^2/v_F^2)^{-1/2}.
\end{equation}

\noindent Where $\gamma$ is the QLT factor and the subscript ``$0$" represents measurements in the moving frame with velocity $u$ (following the notation of Ref.~\cite{Dong2021FizeauPlasmonics}).

The perturbed density is thus transformed as

\begin{equation} \label{eq:QLT-density}
    f^{u}(\epsilon) = \frac{1}{e^{(\epsilon- \mathbf{u} \cdot \mathbf{p} - \mu^u )/k_BT}+1} = \frac{1}{e^{(\epsilon_0 -\mu_0)/\gamma k_BT}+1},
\end{equation}

\noindent where we have introduced $\mu_0 =\gamma \mu^u$. The chemical potential must be transformed such that the quasiparticle density, $n$, is conserved: $n_0 = n/\gamma$ ~\cite{Dong2021FizeauPlasmonics}.
The quasiparticle density of a WSM/DSM is $n = \int_0^{\epsilon_F} N(\epsilon) d \epsilon $, where the electron density of states is $N(\epsilon) = \epsilon^2/({2\pi^2\hbar^3v_F^3})$. At temperature of 0K we can show that the quasiparticle density is $n =\epsilon_F^3/(6\pi^2\hbar^3v_F^3) $, and thus the carrier density and chemical potential follow the relationship $n\propto \mu^3$~\cite{Hofmann2015PlasmonLiquids}. Therefore, to conserve carrier density the chemical potential must transform as $\mu_0 = \mu/\gamma^{1/3}$ and subsequently $\mu^u = \mu_0/\gamma = \mu/\gamma^{4/3}$.

 Next, note that according to our definition~\cite{Borgnia2015Quasi-RelativisticGraphene}

\begin{equation}
    \frac{\partial f^{u}(\epsilon)}{\partial \epsilon} = \gamma \frac{\partial f^{0}(\epsilon_0)}{\partial \epsilon_0} .
 \end{equation}

 Additionally, the transformation of $\delta(\epsilon - v|p|)$ is found by making use of the delta function identity from quantum field theory~\cite{Peskin1995AnTheory.}, which is defined by using an arbitrary function $g(x)$ as follows:

\begin{equation}
     \delta(g(x)-g(x_0)) = \frac{1}{|g'(x_0)|} \delta(x-x_0).
 \end{equation}
 
\noindent We express the QLT of the energy, $\epsilon_0$, as 
 
 \begin{equation}
     \begin{aligned}
         g(x)  &= \gamma (x - u p(x)), \\
         g'(x)  &= \gamma (1 - u \frac{dp(x)}{dx}),
     \end{aligned}
 \end{equation}

\noindent where $g(x) = \epsilon_0$ and $x = \epsilon$, and via $\epsilon = v|p|$ we know that $p(x) = \epsilon/v = x/v$. Using the delta function identity, we find
  \begin{equation}
  \begin{aligned}
        \delta(\epsilon_0 - v|p_0|) 
      &= \frac{1}{|\gamma (1 - u/v)|} \delta(\epsilon - v|p|), \\
      &  =\frac{\epsilon_0}{\epsilon} \delta(\epsilon - v|p|). 
  \end{aligned}
 \end{equation}
 
\noindent  Thus,
 \begin{equation} \label{eq:pol2}
\begin{aligned}
    \Pi^{u}(\mathbf{ q}, \omega) 
    &= - \frac{g}{\mathcal{V}} \sum_\mathbf{ p} \int d\epsilon
    \frac{\epsilon}{\epsilon_0} \delta(\epsilon_0 - v|p_0|) \frac{ v^2\mathbf{ q \cdot p}- \epsilon\mathbf{ q \cdot u} } {\epsilon\omega - v^2\mathbf{ p}\cdot \mathbf{q} } \gamma \frac{\partial f^{0}(\epsilon_0)}{\partial \epsilon_0}.
\end{aligned}
\end{equation}

\noindent  Under the QLT $v^2 p^z - \epsilon u = v^2 p_{0}^z/\gamma$ and  $ \epsilon\omega -  v_F^2 \mathbf{ p \cdot q} = \epsilon_0\omega_0 - v_F^2 \mathbf{ p_0 \cdot q_0}$ ~\cite{Borgnia2015Quasi-RelativisticGraphene}:
\begin{equation}
\begin{aligned}
 \epsilon_0\omega_0 - v_F^2 \mathbf{ p_0 \cdot q_0} &= \epsilon_0\omega_0 - v_F^2 p^z q^z
 \\&=   \gamma^2((\omega -u q_z ) \left(\epsilon(p)-up^z\ \right) - v_F^2 \left(p^z-\frac{u}{v_F^2}\epsilon(p)\right) (q^z-\frac{u}{v_F^2}\omega) ) \\
=& \frac{1}{1-u^2/v_F^2} (\omega \epsilon + u^2 qp - \frac{p q}{v_F^2} - \frac{u^2 \epsilon \omega}{v_F^2}) \\
=& \frac{1}{1-u^2/v_F^2} \epsilon \omega (1-u^2/v_F^2) - v_F^2 pq (1-u^2/v_F^2) \\
=& \epsilon \omega - v_F^2 p^zq^z 
\end{aligned}
\end{equation}

To extract the longitudinal conductivity, $\Pi^u_{\rho_z,\rho_z}({q}, \omega)=\Pi^u({q}, \omega)$, we again consider the longitudinal case of $\mathbf{ E} || \mathbf{ q} || \mathbf{\hat{z}}$: 
\begin{equation} \label{eq:pol4}
\begin{aligned}
    \Pi^{u}({q}, \omega) 
    &= - \frac{g}{\mathcal{V}} \sum_\mathbf{p_0} \int d\epsilon_0
    \frac{\epsilon}{\epsilon_0} \delta(\epsilon_0 - v|p_0|) \frac{ q v_F^2 p_0^z/\gamma} {\epsilon_0\omega_0 - p_0^z v_F^2 q_0} \gamma \frac{\partial f^{0}(\epsilon_0)}{\partial \epsilon_0}.
\end{aligned}
\end{equation}

\noindent Solving the $\epsilon_0$ integral over the delta function we get $\epsilon_0 = v|\mathbf{ p_0}|$, leading to

  \begin{equation} \label{eq:pol5}
\begin{aligned}
    \Pi^{u}({q}, \omega) & =  
    & =   - \frac{g}{\mathcal{V}} \sum_\mathbf{p_0}
    \frac{\epsilon}{\epsilon_0} \frac{ q v_F^2 p_0^z } {\epsilon_0\omega_0 - p_0^z v_F^2 q_0} \frac{\partial f^{0}(\epsilon_0)}{\partial \epsilon_0} . 
\end{aligned}
\end{equation}

\noindent Note that $ \frac{\epsilon}{\epsilon_0} = \gamma (1+ u p_0^z/\epsilon_0)$, and thus
 
\begin{equation} \label{eq:Fizeau-L-v-NL}
\begin{aligned}
    \Pi^{u}({q}, \omega) & =  
    &= - \frac{g}{\mathcal{V}} \sum_\mathbf{p_0} \gamma (1+ \frac{u p_0^z}{\epsilon_0}) \frac{ q v_F^2 p_0^z } {\epsilon_0\omega_0 - p_0^z v_F^2 q_0}\frac{\partial f^{0}(\epsilon_0)}{\partial \epsilon_0} .
\end{aligned}
\end{equation}

\noindent The integral can then be transformed from 3D $p_0$-space to over a constant energy surface $\epsilon_0$ via Eq.~\ref{eq:solid-angle-transform}~\cite{Coleman2015Finite-temperaturePhysics, Borgnia2015Quasi-RelativisticGraphene}

\begin{equation}
\begin{aligned}
    \Pi^{u}({ q}, \omega)  
    & = - g\int_0^{2\pi} d\phi \int_0^{\pi} \frac{d\theta sin\theta}{4\pi} \int d\epsilon_0 N(\epsilon_0)~ 
    \gamma \left(1+ \frac{u p_{0}\cos{\theta}}{\epsilon_0} \right) \frac{ q v_F^2 p_{0}\cos{\theta} } {\epsilon_0\omega_0 - p_{0}\cos{\theta} v_F^2 q_0}
    \frac{\partial f^{0}(\epsilon_0)}{\partial \epsilon_0}.
\end{aligned}
\end{equation}

\noindent  This can be evaluated for low temperature $kT \ll \mu$ using ${\partial f^{0}(\epsilon_0)}/{\partial \epsilon_0} \approx - \delta \left( \epsilon_0 - \mu_0 \right) $ ~\cite{Peskin1995AnTheory.,Borgnia2015Quasi-RelativisticGraphene}:

\begin{equation}
\begin{aligned}
    \Pi^{u}({ q}, \omega) 
    & = g N(\mu_0) \int_0^{2\pi} d\phi \int_0^{\pi} \frac{d\theta sin\theta}{4\pi} ~ \gamma \left(1+ \frac{u p_{0}\cos{\theta}}{\mu_0} \right) \frac{ q v_F^2 p_{0}\cos{\theta} } {\mu_0\omega_0 - p_{0}\cos{\theta} v_F^2 q_0}.
\end{aligned}
\end{equation}

\noindent Note that $\frac{1}{v_F} = \frac{p_{0}(\mu_0)}{\mu_0}$, and thus

\begin{equation}
\begin{aligned}
    \Pi^{u}({ q}, \omega) & = g N(\mu_0) 2\pi \int_0^{\pi} \frac{d\theta sin\theta}{4\pi} ~ 
    \gamma \left(1+ \frac{u\cos{\theta}}{v_F} \right) \frac{ v_F q\cos{\theta} } {\omega_0 - v_F q_0 \cos{\theta} }. 
\end{aligned}
\end{equation}

\noindent Let $a=\cos\theta$ , $\frac{da}{d\theta} = -\sin\theta$, and $d\theta \sin\theta = -da$:
\begin{equation}
\begin{aligned}
    \Pi^{u}({q}, \omega) & = g N(\mu_0) \frac{\gamma}{2} \int_1^{-1} -da ~ 
    \left( 1+ \frac{u a}{v_F} \right) \frac{ v_F q a } {\omega_0 - v_F q_0 a }, \\
     & = -g N(\mu_0) \frac{1}{2} 
     \frac{q \gamma (q_0 v_F^2 +u\omega_0)}{q_0^3 v_F^3} 
     \left( \omega_0 \ln \left( \frac{\omega_0 - q_0 v_F}{\omega_0 + q_0 v_F}\right) +2 q_0 v_F
    \right), \\
    & = - g N(\mu_0) 
     \frac{q^2}{q_0^2} 
     \left( \frac{\omega_0}{2 q_0 v_F} \ln \left( \frac{\omega_0 - q_0 v_F}{\omega_0 + q_0 v_F}\right) +1
    \right) .
\end{aligned}
\end{equation}

\noindent The longitudinal polarizability is converted to longitudinal optical conductivity as follows:

\begin{equation} \label{eq:Fizeau-cond}
\begin{aligned}
    {\mbox{Im}\sigma^u_{zz}} (q,\omega) & = \frac{ie^2\omega}{q^2} \mbox{Re}\Pi^u (q, \omega), \\ 
    & = -\frac{ie^2\omega}{q_0^2} g N(\mu_0) 
     \left( \frac{\omega_0}{2 q_0 v_F} \ln \left( \frac{\omega_0 - q_0 v_F}{\omega_0 + q_0 v_F}\right) +1
    \right) . \\ 
\end{aligned}
\end{equation}
\noindent Comparing Eq.~\ref{eq:Fizeau-cond} to Eq.~\ref{eq:Normal-cond-Im}, we have shown that the longitudinal optical conductivity is transformed under Fizeau drag as:

\begin{equation} \label{eq:Fizeau-transform-result}
\begin{aligned} 
        \sigma_{zz}^{u} (\omega,q,\mu)
        &=  \frac{\omega}{\omega_0}\sigma_{zz}(\omega_0,q_0,\mu_0), \\
\end{aligned}
\end{equation}
where
\begin{equation} \label{eq:QLTs}
\begin{aligned} 
        \omega_0=\gamma(\omega -u q_z ),~
        q_{z,0}&=\gamma(q_z-\frac{u}{v_F^2}\omega),~
        q_{x,0}=q_x,~
        q_{y,0}=q_y, ~
        \mu_0 = \frac{\mu}{\gamma^{1/3}}.
\end{aligned}
\end{equation}

\subsection*{Local case }
For thoroughness and comparison, we show that the same Fizeau drag derivation can be performed for the local case, making use of Taylor series expansion around $q_0 \rightarrow 0$ starting from equation~\ref{eq:Fizeau-L-v-NL}, the middle factor becomes:

\begin{equation} \label{eq:polar-xx}
\begin{aligned}
    \frac{ q v_F^2 p_0^z} {\epsilon_0\omega_0 - p_0^z v_F^2 q_0} = \frac{ q v_F^2 p_0^z} {\epsilon_0\omega_0} + \frac{ qq_0 v_F^4 (p_0^z)^2} {\epsilon_0^2\omega_0^2} + O[q]^2, \\ 
\end{aligned}
\end{equation}

\noindent which multiplies with the term in parenthesis:

\begin{equation} \label{eq:polar-xx-b}
\begin{aligned}
    (1+ \frac{u p_0^z}{\epsilon_0})\frac{ q v_F^2 p_0^z} {\epsilon_0\omega_0 - p^x_0 v_F^2 q_0} &\approx 
    (1+ \frac{u p_0^z}{\epsilon_0}) \left( \frac{ q v_F^2 p_0^z} {\epsilon_0\omega_0} + \frac{ qq_0 v_F^4 (p_0^z)^2} {\epsilon_0^2 \omega_0^2} \right), \\
    &\approx \frac{ q u v_F^2 (p_0^z)^2} {\epsilon^2_0 \omega_0} + \frac{ q q_0 v_F^4 (p_0^z)^2} {\epsilon_0^2 \omega_0^2}, \\ 
     &= \frac{ q^2 v_F^4 (p_0^z)^2}{\epsilon^2_0 \omega^2_0} \frac{1}{\gamma} . \\ 
\end{aligned}
\end{equation}
\noindent  Thus, we arrive at
 \begin{equation} \label{eq:pol11}
\begin{aligned}
    \Pi^u({q}, \omega) & \approx  - \frac{g}{\mathcal{V}} \sum_\mathbf{p_0} ~  
    \frac{ q^2 v_F^4 (p_0^z)^2}{\epsilon^2_0\omega^2_0} \frac{\partial f^{0}(\epsilon_0)}{\partial \epsilon_0} .
\end{aligned}
\end{equation}
\noindent  The integral can then be transformed from 3-dimensional $p$-space to over a constant energy surface via Eq.~\ref{eq:solid-angle-transform} 

\begin{equation} \label{eq:pol8}
\begin{aligned}
    \Pi^u({ q}, \omega) & = - g\int_0^{2\pi} d\phi \int_0^{\pi} \frac{d\theta sin\theta}{4\pi} \int d\epsilon N(\epsilon)~  
    \frac{ q^2 v_F^4 (p_{0}(\epsilon_0)\cos{\theta})^2}{\epsilon^2_0\omega^2_0} \frac{\partial f^{0}(\epsilon_0)}{\partial \epsilon_0} .
\end{aligned}
\end{equation}

\noindent  As for the nonlocal case, this is evaluated for low-temperature $kT \ll \mu$ using ${\partial f^{0}(\epsilon_0)}/{\partial \epsilon_0} \approx - \delta \left( \epsilon_0 - \mu_0 \right) $

\begin{equation} \label{eq:pol10}
\begin{aligned}
    \Pi^u({q}, \omega) & =  g\frac{ q^2 v_F^2 }{\omega^2_0} N(\mu_0)\int_0^{2\pi} d\phi \int_0^{\pi} \frac{d\theta }{4\pi}\sin{\theta}\cos^2{\theta} , \\
    & =   \frac{1}{3} N(\mu_0)g  
    \frac{ q^2 v_F^2 }{\omega^2_0} ,  \\
    & = \frac{1}{3} \frac{g\mu_0^2}{2\pi^2\hbar^3v_F}~  
    \frac{ q^2}{\omega^2_0} .
\end{aligned}
\end{equation}

\noindent The longitudinal polarizability is converted to longitudinal optical conductivity 

\begin{equation}
\begin{aligned}
    \sigma^u_{zz}(q,\omega) &= \frac{\omega}{\omega_0}  \frac{i e^2 g \mu_0^2}{6\pi^2\hbar^3\omega_0v_F}, \\
    &= \frac{\omega}{\gamma(\omega-uq)}  \frac{i e^2 g \mu_0^2}{6\pi^2\hbar^3v_F\cdot\gamma(\omega-uq)}.
\end{aligned}   
\end{equation}

\noindent When compared to the non-current carrying case, it is clear that the longitudinal optical conductivity is transformed again following equations ~\ref{eq:Fizeau-transform-result} and~\ref{eq:QLTs}. 

\newpage
\section*{Fizeau Drag in RPA vs. local Drude vs. nonlocal Drude models}
\begin{figure} [H]
    \centering
    \includegraphics[width=1\linewidth]{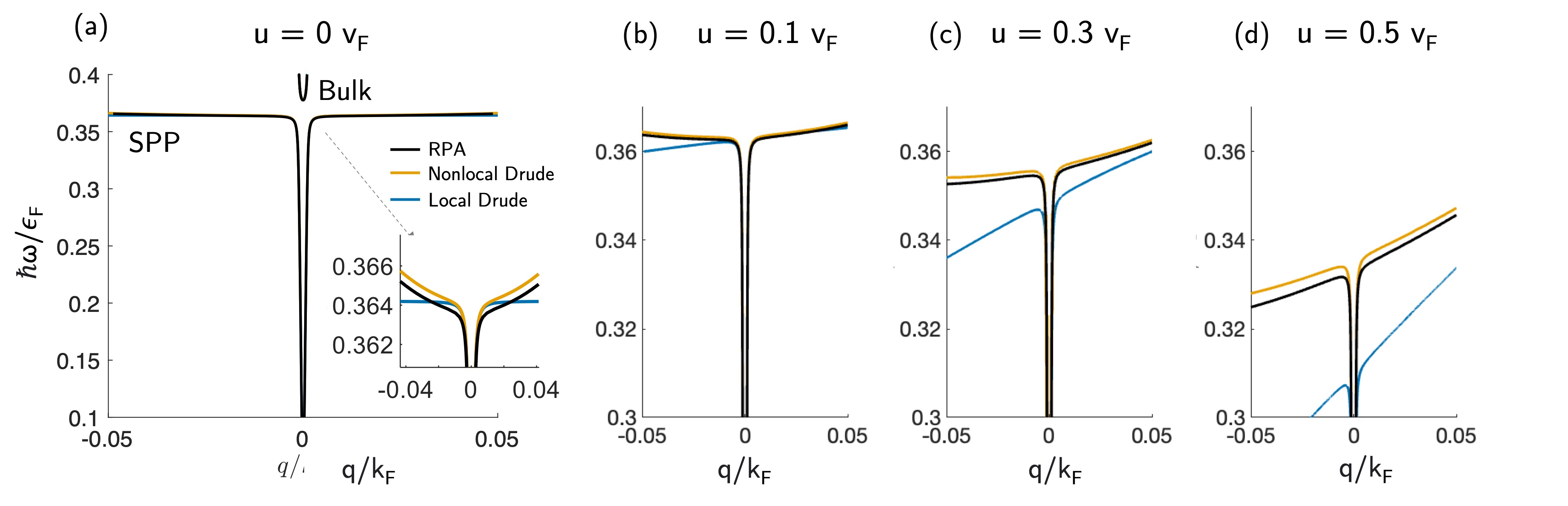}
    \caption{The lossless SPP dispersion for Cd$_3$As$_2$ ($v_F=1\times10^{6}$m/s, $\epsilon_F = 0.15$ eV, and $g=4$) for the local Drude~\cite{Kotov2016DielectricFilms} (Eq.~\ref{eq:pol-SC-local}), nonlocal Drude (Eq.~\ref{eq:pol-normal}), and the RPA dielectric model presented in Ref.~\cite{Thakur2017DynamicalSystems,Thakur2018DynamicSemimetals} (Eq.~\ref{eq:pol-RPA}), for the (a) unbiased case and (b) $u=0.1v_F$, (c) $u=0.3v_F$, and (d) $u=0.5v_F$. }
    \label{fig:NL-L_RPA-SPP-comparisons}
\end{figure}

In Suppl. Fig.~\ref{fig:NL-L_RPA-SPP-comparisons}(a-d) we plot the SPP dispersion for lossless Cd$_3$As$_2$ for increasing current biases using the local Drude, Eq.~\ref{eq:pol-SC-local}, nonlocal Drude, Eq.~\ref{eq:pol-normal}, and RPA, Eq.~\ref{eq:pol-RPA}, dielectric models. The RPA and nonlocal Drude models closely match, while the local Drude model does not align with either result. With increased carrier velocity, the local and nonlocal optical models increasingly diverge, resulting in vastly different predicted SPP frequencies, where the local model overestimates the magnitude of Fizeau shift. This emphasizes the importance of accounting for nonlocality in Fizeau drag calculations.

The nonlocal predictions for Fizeau drag differs notably from the prediction of truly unidirectional SPP transport which can be made when invoking the non-realistic local Drude mode. The local Drude model predicts that the counter-propagating mode will never bend upward to compete with the co-propagating mode, facilitating unidirectionality. 

\section*{Fizeau drag at large q-values}

\begin{figure} [ht]
    \centering
    \includegraphics[width=0.5\linewidth]{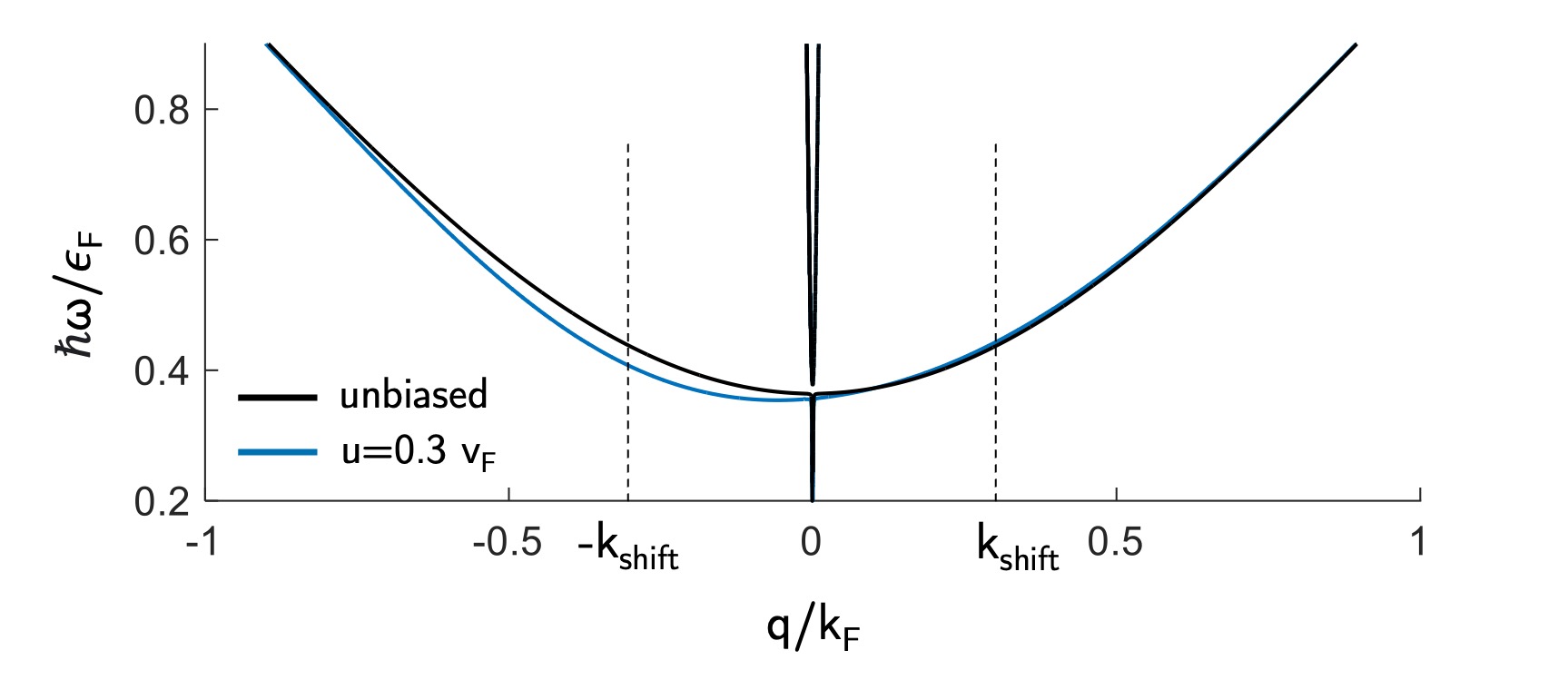}
    \caption{As $q \rightarrow k_{shift}$ and beyond, the dispersion plots of the SPP mode and the higher energy mode (blue), which had shifted due to the current bias, re-converge with their unbiased counterparts (black). SPP modes are shown for $u=0.3v_F$ in Cd$_3$As$_2$ using the same parameters as in Fig.~\ref{fig:fizeauResults} in the manuscript.}
    \label{fig:q-to-kshift}
\end{figure}

Finally, we show in Fig.~\ref{fig:q-to-kshift} the SPP dispersion plot for a larger range of q-values to verify that for q values much larger than the shift of the Fermi disk, $k_{shift}= k_F  sgn[\epsilon_F] u/v_F$, the current-biased SPP dispersion converges with the normal, un-biased SPP dispersion plot.

\newpage
\section*{Supplemental Note 4: Complex-q root search method. The role of collision-driven dissipative loss and Landau damping}\label{sec:A3}

\begin{figure} [ht]
    \centering
    \includegraphics[width=0.7\linewidth]{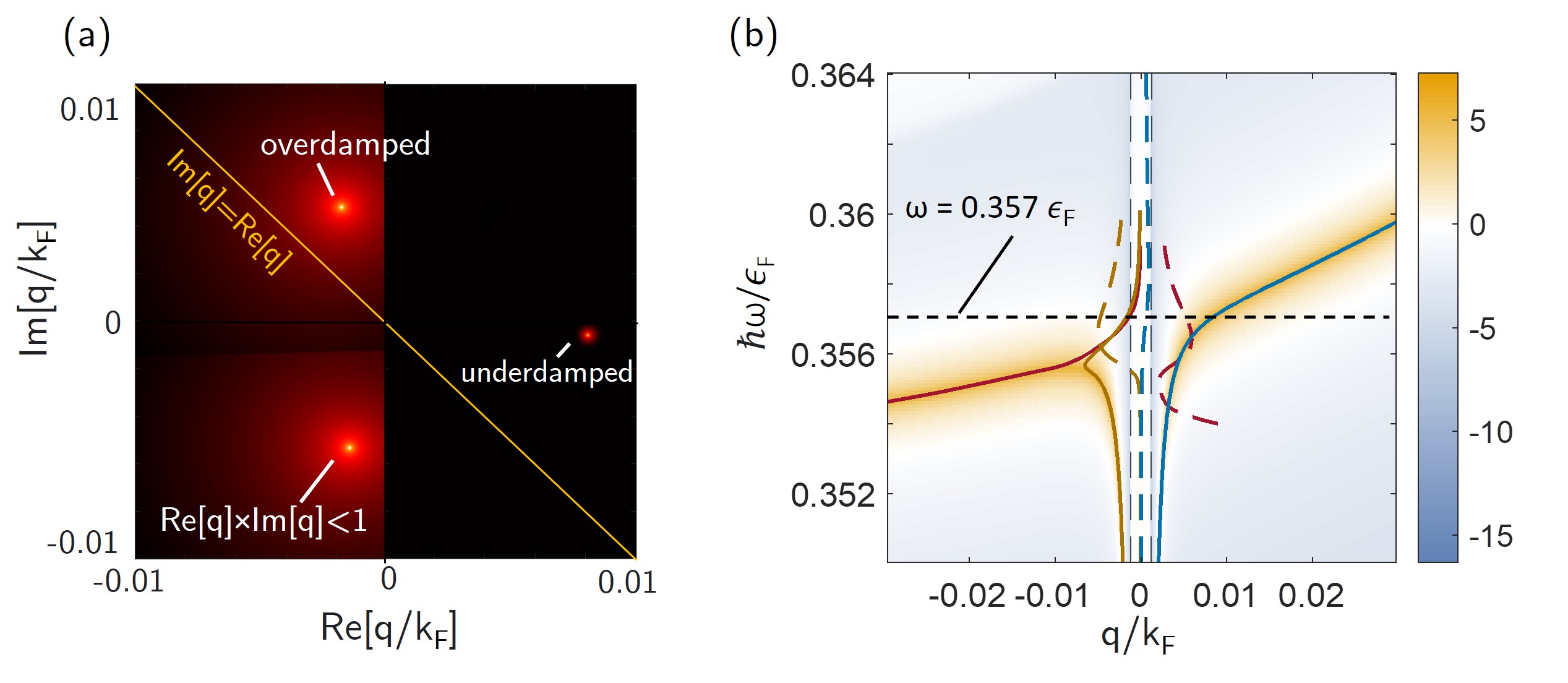}
    \caption{\textbf{Complex q-space solution for increasing loss. } (a) The equifrequency complex wavevector solution space for $\omega =0.357\epsilon_F$ for (a) $u=0.3v_F$, $\tau = 2.1\times 10^{-11}s$ and (b) the corresponding SPP dispersion. The solid line corresponds to the $\mbox{Re}(q)$ value and the dashed line shows the $\mbox{Im}(q)$ value. The red line corresponds to the unphysical mode exhibiting gain, $\mbox{Re}(q)/\mbox{Im}(q)>1$. The same Cd$_3$As$_2$ parameters are used as in Fig.~\ref{fig:lossy-result}.}
    \label{fig:complex-q-space}
\end{figure}

All the SPP mode solutions shown in this manuscript have been found by searching over a complex $q = \mbox{Re}[q] + \mbox{Im}[q]$ plane for the roots of the corresponding dispersion equation (see Suppl. Fig.~\ref{fig:complex-q-space}(a,b)). The solutions corresponding to lossy modes, with $\mbox{Re}[q]\cdot\mbox{Im}[q]>1$, are included in the main  figures of the manuscript. In turn, Fig.~\ref{fig:complex-q-space}(a) also shows an overdamped ($|\mbox{Im}[q]|>|\mbox{Re}[q]|$) and underdamped ($|\mbox{Im}[q]|<|\mbox{Re}[q]|$) modes. The complex dispersion relation has an additional spurious solution corresponding to a -q propagating SPP mode, which is characterized by $\mbox{Re}[q]\cdot\mbox{Im}[q]<1$, signifying a gain mode, included in Fig.~\ref{fig:complex-q-space} in red.
This gain mode is visible in the $\mbox{Im} r_{pp}$ plots in the main text. 

\begin{figure} [ht]
    \centering
    \includegraphics[width=0.4\linewidth]{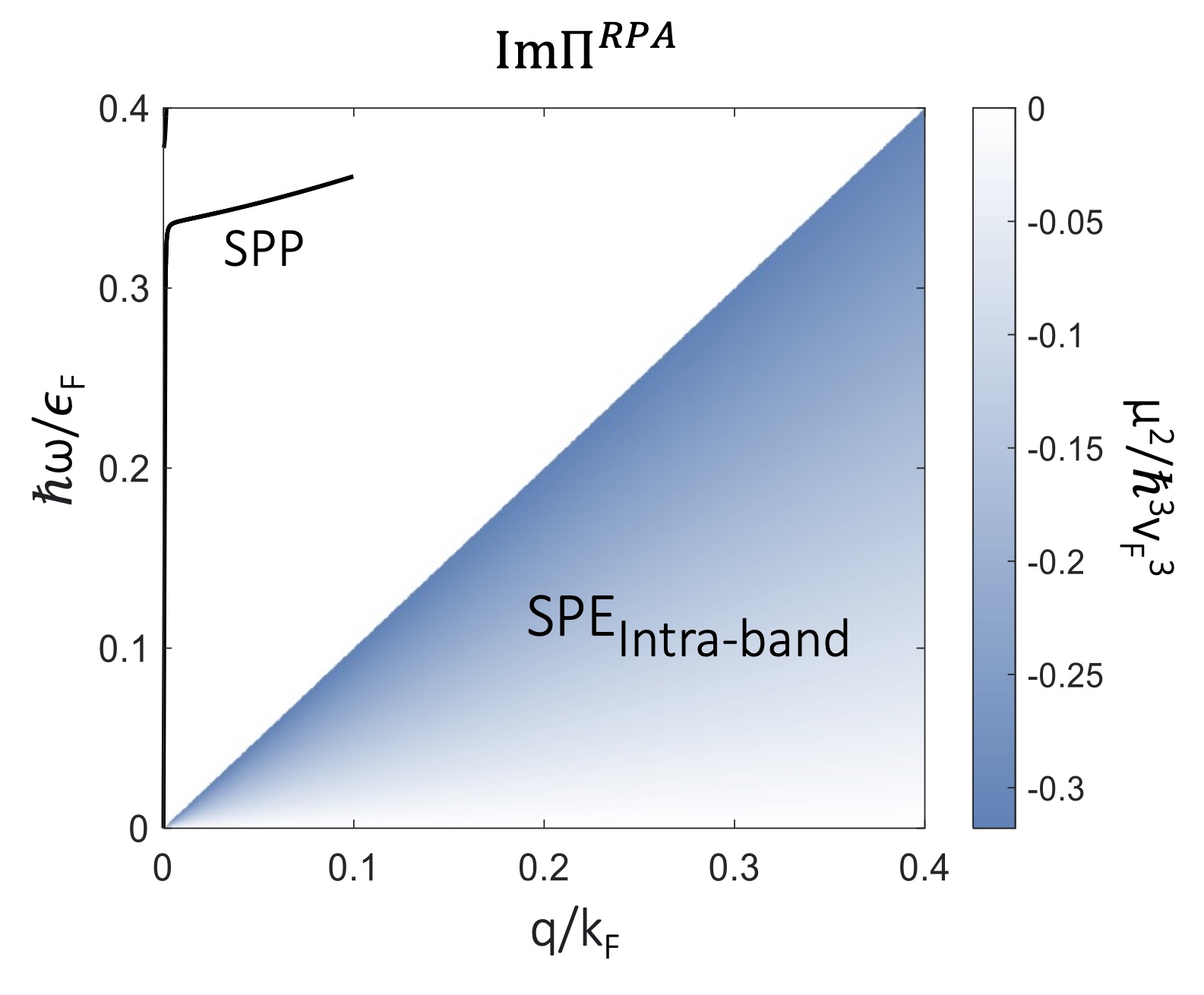}
    \caption{Plot of $\mbox{Im}\Pi_{\rho,\rho}$ in units of $\mu^2 /\hbar^3 v_F^3$ shows that the single-particle excitation (SPE) continuum where the intra-band transitions are allowed is bounded by $\omega = v_F q$~\cite{Lv_Zhang_2013}. Note that the SPE region corresponding to the inter-band transitions is at higher energies not shown on the plot. The SPP mode corresponding to that in Fig.~\ref{fig:lossy-result} is shown.}
    \label{fig:SPE_continuum}
\end{figure}

It should be noted that plasmon gain originating from inter-band transitions has been proposed in current biased tilted-node Dirac semimetals~\cite{Park2022PlasmonicNodes}. 
In our material system, the SPP modes are well separated from the regions of both the intra-band and the inter-band transitions in the energy-momentum space. In particular, Landau damping from electron-hole excitations occurs at momentum values exceeding the axes in Fig.~\ref{fig:lossy-result}(a-b). 
This is evident from the examination of the plot of $\mbox{Im}~r_{pp}$ in Fig.~\ref{fig:SPE_continuum}, which shows that the single-particle excitation continuum is bounded by $\omega = v_Fq$~\cite{Lv_Zhang_2013}.
Furthermore, inter-band transitions are not included into our semiclassical nonlocal current-biased model of dielectric permittivity, which rules out any possible physical mechanisms for the SPP mode amplification. Finally, as discussed in the prior work on current-biased SPP modes propagating along the interfaces of noble metals~\cite{Gangaraj2022DriftingPhotonics}, these modes cannot exhibit gain via negative Landau damping because a single medium with current cannot cause mode amplification (see Ref.~\cite{Morgado2017NegativeGraphene} for detail). 
Thus, in our work, we neglect this gain mode as unphysical. 
Conversely, we conclude that the +q mode is preserved until the Landau damping momentum threshold is reached, which occurs outside the q range shown in  Fig.~\ref{fig:lossy-result}. 

\begin{figure} [ht]
    \centering
    \includegraphics[width=1\linewidth]{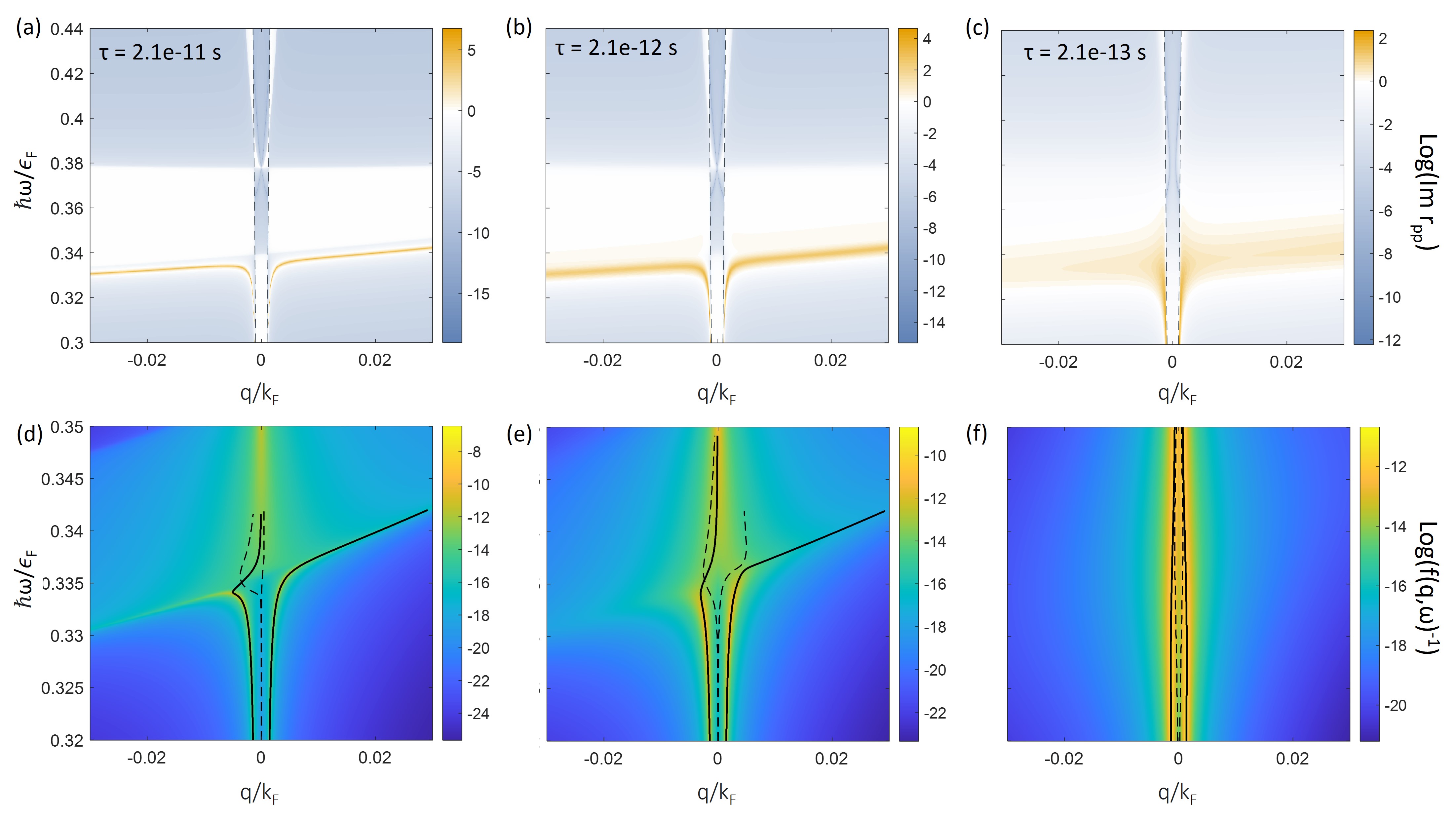}
    \caption{The loss functions for the semi-infinite current-biased Cd$_3$As$_2$-air interface represented as either (a-c) 
    $L(q,\omega)= \mbox{Im}~r_{pp}$ or (d-f) $L(q,\omega)=\mbox{Im}(f(q,\omega)^{-1})$ for a varying relaxation time (a \& d) $\tau=2.1\times10^{-11}$s, (b \& e) $\tau=2.1\times10^{-12}$s, and (c \& f) $\tau=2.1\times10^{-13}$s. The Cd$_3$As$_2$ is modeled under current bias of drift velocity $u=0.5v_F$ with the same parameters as Fig.~\ref{fig:fizeauResults}. Plots (d-f) show an overlay of the solutions of the complex dispersion relation on top of the loss function heat map.}
    \label{fig:lossFunctionComparisons}
\end{figure}

Overall, our modeling shows that as the dissipative losses increase, the $\mbox{Log}(\mbox{Im}~r_{pp})$ loss function peaks lose their strength, indicating that the SPP modes become harder to excite. 
Fig.~\ref{fig:lossFunctionComparisons} shows that at the loss level quantified by $\tau=2.1\times10^{-13}$s, SPP mode overdamping is evident in both loss function plots and in the plots of the complex solutions of the dispersion equation. Note that the loss function is typically represented for the surface modes on interfaces of 3D materials as either $L(q,\omega)=\mbox{Im}~ r_{pp}$~\cite{Costa_Gonçalves_Basov_Koppens_Mortensen_Peres_2021} or $L(q,\omega)=\mbox{Im} (-f(q,\omega)^{-1})$, where $f(q,\omega)=0$ is the SPP dispersion relation~\cite{Kotov2016DielectricFilms}. For this reason, we plot both loss functions for comparison in Fig.~\ref{fig:lossFunctionComparisons}. 

Finally, we note that plotting the $\mbox{Log}(\mbox{Im}~r_{pp})$ loss function reveals the presence of a continuum of modes at energies above the SPP branches, shown as a white band in Fig.~\ref{fig:lossFunctionComparisons}. This continuum is similar to the bulk continuum present in the dispersion of TRS-breaking WSM SPPs in the Voigt configuration~\cite{Tang_Chen_Zhang_2021, Zhao2020Axion-Field-EnabledSemimetals}, as well as the continuum in energy above the SPPs in graphene on superconductor~\cite{Costa_Gonçalves_Basov_Koppens_Mortensen_Peres_2021} and silver films~\cite{Echarri_Gonçalves_Tserkezis_Abajo_Mortensen_Cox_2021}.

In Suppl. Fig.~S8(a-c) we show the SPP dispersion plots for Cd$_3$As$_2$ with progressively increased levels of dissipative losses, from $\tau=2.1\times 10^{-10}$s to $2.1\times 10^{-12}$s, without any Fizeau drag.

\begin{figure} [H]
    \centering
    \includegraphics[width=1\linewidth]{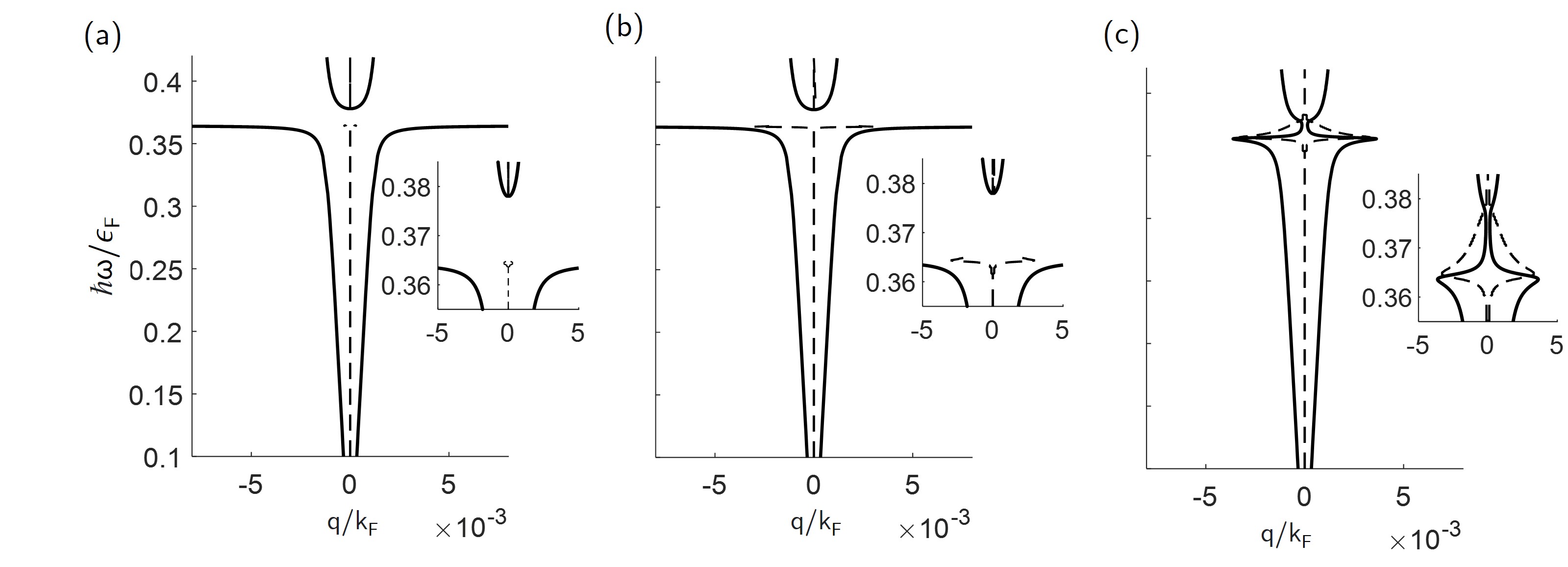}
    \caption{The SPP dispersion for Cd$_3$As$_2$ ($v_F=1\times10^{6}$m/s, $\epsilon_F = 0.15$ eV, and $g=4$) for the nonlocal Drude model derived in Eq.~\ref{eq:pol-normal} without DC-bias and for (a) $\tau=2.1\times 10^{-10}$s, (b) $\tau=2.1\times 10^{-11}$s and (c) $\tau=2.1\times 10^{-12}$s. }
    \label{fig:noFizeaulossy}
\end{figure}

\end{document}